\documentclass[superscriptaddress,twocolumn,notitlepage,pra,longbibliography]{revtex4-2}
\usepackage[colorlinks=true,citecolor=blue,urlcolor=black]{hyperref}
\usepackage{graphicx}
\usepackage{dcolumn}
\usepackage{bm}
\usepackage{amsthm}
\usepackage{amsmath}
\usepackage{amssymb}
\usepackage{color}
\usepackage{multirow}
\usepackage{algorithm}
\usepackage{algorithmicx}
\usepackage{algpseudocode}
\usepackage{subfigure}
\usepackage{hyperref}
\usepackage{wasysym}
\usepackage{xcolor}

\definecolor{Red}{rgb}{1,0,0}
\def\vec#1{{\bm #1}}

\def\ket#1{| #1 \rangle}
\def\bra#1{\langle #1 |}

\def\Tr{\operatorname{Tr}}

\begin{document}


\title{Robust resource-efficient quantum variational ansatz through evolutionary algorithm}

\author{Yuhan Huang}
\affiliation{Institute of Fundamental and Frontier Sciences, University of Electronic Science and Technology of China, Chengdu, 610051, China}

\author{Qingyu Li}
\affiliation{Institute of Fundamental and Frontier Sciences, University of Electronic Science and Technology of China, Chengdu, 610051, China}

\author{Xiaokai Hou}
\affiliation{Institute of Fundamental and Frontier Sciences, University of Electronic Science and Technology of China, Chengdu, 610051, China}

\author{Rebing Wu}
\affiliation{Department of Automation, Tsinghua University, Beijing 100084, P. R. China}

\author{Man-Hong Yung}
\affiliation{Central Research Institute, 2012 Labs, Huawei Technologies}

\author{Abolfazl Bayat}
\email{abolfazl.bayat@uestc.edu.cn}
\affiliation{Institute of Fundamental and Frontier Sciences, University of Electronic Science and Technology of China, Chengdu, 610051, China}

\author{Xiaoting Wang}
\email{xiaoting@uestc.edu.cn}
\affiliation{Institute of Fundamental and Frontier Sciences, University of Electronic Science and Technology of China, Chengdu, 610051, China}

\begin{abstract}

Variational quantum algorithms (VQAs) are promising methods to demonstrate quantum advantage on near-term devices as the required resources are divided between a quantum simulator and a classical optimizer. As such, designing a VQA which is resource-efficient and robust against noise is a key factor to achieve potential advantage with the existing noisy quantum simulators. It turns out that a fixed VQA circuit design, such as the widely-used hardware efficient ansatz, is not necessarily robust against imperfections. In this work, we propose a genome-length-adjustable evolutionary algorithm to design a robust VQA circuit that is optimized over variations of both circuit ansatz and gate parameters, without any prior assumptions on circuit structure or depth. Remarkably, our method not only generates a noise-effect-minimized circuit with shallow depth, but also accelerates the classical optimization by substantially reducing the number of parameters. In this regard, the optimized circuit is far more resource-efficient with respect to both quantum and classical resources. As applications, based on two typical error models in VQA, we apply our method to calculate the ground energy of the hydrogen and the water molecules as well as the Heisenberg model. Simulations suggest that compared with conventional hardware efficient ansatz, our circuit-structure-tunable method can generate circuits apparently more robust against both coherent and incoherent noise, and hence is more likely to be implemented on near-term devices.

\end{abstract}

\maketitle

\section{Introduction}

Thanks to recent technological advances, quantum devices are rapidly scaling up in size with improved performance, significantly increasing the possibility of realizing true quantum simulations~\cite{lloyd1996universal,buluta2009quantum,georgescu2014quantum,altman2021quantum}. Nonetheless, achieving fault-tolerant quantum computing remains a huge challenge. Current noisy intermediate-scale quantum (NISQ) devices are far from being perfect in their initialization, operation and readout~\cite{preskill2018quantum}. Therefore, the quest for achieving quantum advantage with NISQ devices is heating up~\cite{bharti2021noisy}. Variational quantum algorithms~\cite{cerezo2021variational,jones2019variational}, are among the most promising approaches for achieving quantum advantage in NISQ era. In such algorithms, the complexity of the system is divided between a quantum simulator and a classical optimizer, with the hope that an imperfect shallow NISQ circuit will reach a quantum advantage. The quantum variational algorithms have been found useful for several applications in various fields, including computational chemistry~\cite{o2016scalable,kandala2017hardware,colless2018computation,mccaskey2019quantum,mcardle2020quantum,arute2020hartree}, simulating strongly correlated systems~\cite{kokail2019self,lyu2020accelerated,lau2021quantum,haug2020generalized} and their phase detection~\cite{gong2022quantum}, optimization~\cite{farhi2014quantum,farhi2016quantum,wang2018quantum,moll2018quantum,harrigan2021quantum}, solving linear~\cite{harrow2009quantum,bravo2019variational,xu2021variational} and nonlinear equations~\cite{lubasch2020variational}, classification problems~\cite{schuld2019quantum,havlivcek2019supervised}, generative models~\cite{dallaire2018quantum,benedetti2019generative,du2020expressive} and quantum neural networks~\cite{hou2021universal,farhi2018classification}. Among these algorithms, the variational quantum eigensolver (VQE)~\cite{peruzzo2014variational,aspuru2005simulated,kandala2017hardware}, as a special type of VQAs, has been developed for efficiently generating the ground state of many-body systems on quantum simulators. The VQE has been widely used in quantum chemistry~\cite{o2016scalable,kandala2017hardware,colless2018computation,mccaskey2019quantum,arute2020hartree,mcardle2020quantum,peruzzo2014variational,google2020hartree} and condensed matter physics~\cite{kokail2019self,lyu2020accelerated,lau2021quantum,haug2020generalized}. It has also been generalized to simulate higher-energy eigenstates~\cite{higgott2019variational,mcclean2017hybrid,nakanishi2019subspace,santagati2018witnessing}, time evolution~\cite{li2017efficient,yuan2019theory}, Gibbs thermal states~\cite{chowdhury2020variational,wang2021variational} and non-equilibrium steady states~\cite{yoshioka2020variational} of many-body systems. There are several realizations of VQE in photonic chips~\cite{peruzzo2014variational}, ion traps~\cite{hempel2018quantum,kokail2019self,shen2017quantum}, nuclear magnetic resonance systems~\cite{li2011solving}, and superconducting
quantum devices~\cite{o2016scalable,kandala2017hardware,colless2018computation,arute2020hartree}.

From a resource perspective, all quantum variational algorithms, including the VQE, demand two types of resources: (i) a classical resource which is quantified through the number of parameters and the number of iterations required for minimizing the average energy; and (ii) a quantum resource which is quantified through the depth of the circuit. 	Fulfilling a quantum variational algorithm is indeed very time-consuming as it demands several iterations with many repetitions of experiments. It is highly important to make them more efficient with respect to both quantum and classical resources. Several proposals have been dedicated to accelerate the classical optimization through variations of gradient descent~\cite{sweke2020stochastic,izmaylov2021analytic}, proper initial guess for the circuit parameters~\cite{lyu2020accelerated} and sequential minimal optimization~\cite{nakanishi2020sequential}. 
The situation is more complex for quantum resources. In order to simplify a quantum circuit one can exploit symmetries~\cite{gard2020efficient,Lyu2022symmetry} and canonical transformations~\cite{ryabinkin2018qubit,ryabinkin2020iterative}. However, these methods normally rely on a fixed structure in the design of their quantum circuit. Although, this makes the generalization for larger systems straightforward they may not be very resource-efficient with respect to the circuit depth and also may not be robust against imperfections in the system. Therefore, designing a quantum circuit which is both robust and resource-efficient (i.e. as shallow as possible) is highly desirable.

A key issue in the NISQ era is, in the absence of error correction, how to enhance the robustness of quantum algorithms against various types of imperfections, ranging from distorted control pulses for elementary gate operations, crosstalk between qubits, incoherence due to system-environment interaction, to faulty readout apparatus. It has been shown that error mitigation methods at the readout side are effective to reduce imperfections~\cite{temme2017error,endo2018practical,ai2021exponential,lowe2021unified,o2021error}. Besides, control theory also provides an insightful framework discussing how to achieve robustness against noise. The essential idea is to construct an optimal control that will deal with random errors varying over a certain range~\cite{li2006control,chen2014sampling,wu2019learning,wu2020end,ge2020robust,turinici2019stochastic,kang2021batch,dong2020robust}. For variational quantum algorithms including VQE, there are two types of variations that can be made on quantum circuit: one is the variation of parameters in each elementary gate~\cite{lavrijsen2020classical,lyu2020accelerated,stokes2020quantum,nakanishi2020sequential}, and the other is the variation of the circuit structure~\cite{grimsley2019adaptive,li2020quantum,du2020quantum,cincio2021machine,kuo2021quantum,altares2021automatic}. Conventional discussions of robustness have focused on the optimization of parameters for a fixed circuit ansatz. In the following, we will see that for VQE algorithms, such variation will not improve robustness substantially. In order to solve this problem, we propose in this work to implement both types of variations simultaneously in order to achieve a robust circuit design.


Two general methods have been developed for designing quantum circuits in variational quantum algorithms. In the first approach, independent of the task, a fixed design of a circuit such as the hardware efficient ansatz (HEA)~\cite{kandala2017hardware} is concatenated several times to make the quantum circuit. In the second approach, however, the quantum circuit is designed for a specific task, e.g. a given Hamiltonian with certain symmetries in the VQE algorithm~\cite{gard2020efficient,lyu2020accelerated}. Although, the first approach makes the design of the hardware simpler it may not be the most suitable strategy due to over-sized number of parameters, barren plateau phenomenon~\cite{mcclean2018barren} and more importantly robustness against imperfections. Therefore, it has been an increasing interest in recent years for developing new methods for quantum circuit design. In particular, three different machine learning based approaches have been exploited for designing quantum circuits, including: (i) reinforcement learning methods~\cite{ostaszewski2021reinforcement,pirhooshyaran2020quantum,fosel2021quantum,kuo2021quantum}, (ii) probabilistic learning model~\cite{zhang2020differentiable}, and (iii) evolutionary algorithms~\cite{chivilikhin2020mog,rattew2019domain,wang2021quantumnas,franken2020gradient,du2020quantum,cincio2021machine}. While reinforcement learning methods have been useful in reducing the circuit depth~\cite{ostaszewski2021reinforcement,pirhooshyaran2020quantum,fosel2021quantum,kuo2021quantum}, they suffer from a complicated action space during the training process as well as the difficulty of defining a universal reward function. The probabilistic learning models~\cite{zhang2020differentiable} do not require reward functions, but they have been restricted to quantum circuits with a fixed block structure which severely limits the ansatz domain.
The evolutionary algorithms do not rely on a training process and instead search for the solution among a population which evolves iteratively. This makes them good candidates for optimizing quantum circuits. However, current evolutionary algorithms are also based on a fixed block structure~\cite{du2020quantum} or limited to only mutation operations~\cite{rattew2019domain,franken2020gradient}, which ultimately make their versatility and search domain very restricted. A key open question is whether one can develop a quantum circuit design algorithm which is capable of configuring quantum ansatzes, which are both resource-efficient and robust.

In this work, we introduce two different notions of robustness, namely imperfection- and training-robustness, which are relevant to the VQE algorithm. Then we show that fixed circuit designed ansatzes, such as the widely used HEA, are not robust against imperfections and training. Next we design a Quantum Circuit Evolution of Augmenting Topologies (QCEAT) algorithm, as an evolutionary method for designing quantum circuits. In contrast to previous evolutionary approaches~\cite{chivilikhin2020mog,rattew2019domain}, the QCEAT utilizes several operations such as mutation, intraspecies crossover, and interspecies crossover and is fully flexible in finding the required depth of the circuit without any prior assumption. Remarkably, the QCEAT not only provides circuits which are robust against both imperfection and training but also naturally finds the simple circuit with the minimum number of gates. This not only reduces the circuit depth but also significantly speeds up the classical optimization through minimizing the number of required parameters. 

This paper is organized as follows: in Section II, we review the basic knowledge for VQE; in Section III, we describe two notions of robustness; in Section IV, we introduce the two important models which are used in this paper; in Section V, we study the performance of HEA against different noises for implementing VQE; in Section VI, we present the detail of the QCEAT algorithm inspired by classical genetic algorithm to search quantum circuits; finally, in Section VII, we analyze the robustness of QCEAT-inspired circuits compared with HEA under the different noise models.

\section{Variational Quantum Eigensolver}
In this section, we briefly review the VQE algorithm as a type of hybrid quantum variational method that utilizes a quantum processor and a classical optimizer to generate the ground state $|gs\rangle$ of a given Hamiltonian $H$ through an iterative optimization. At $k$-th iteration, a parameterized quantum circuit which implements a unitary operator $U(\vec\theta^{(k)})$, with $\vec{\theta^{(k)}}{=}(\theta^{(k)}_1,\theta^{(k)}_2,\cdots,\theta^{(k)}_M)$ being the angles of $M$ local 
rotations in the circuit, acts on an initial state $\ket{\phi_0}$ to create the quantum state $U(\vec\theta^{(k)})\ket{\phi_0}$. By performing appropriate measurement at the output of the quantum circuit, one can measure the average energy
\begin{equation}\label{eq:cost_VQE}
E(\vec\theta^{(k)})=\bra{\phi_0} U^\dagger(\vec\theta^{(k)}) H U(\vec\theta^{(k)}) \ket{\phi_0}.
\end{equation}
By feeding $E(\vec\theta^{(k)})$ to a classical optimizer one can minimize the average energy through a gradient-based update rule, such as the gradient descent~\cite{ruder2016overview} or the quasi-Newton~\cite{shanno1970conditioning} methods. In this paper, we use Adam optimizer~\cite{kingma2014adam} as our gradient-based method for minimizing the average energy. The classical optimizer provides an update for the parameters of the circuit at each iteration, namely $\vec\theta^{(k)} \rightarrow \vec\theta^{(k+1)}$, such that $E(\vec\theta^{(k+1)})< E(\vec\theta^{(k)})$. After sufficiently many the number of iterations, the average energy converges to its minimum for an optimal set of parameters $\vec\theta^*$. In this situation, the quantum circuit produces the ground state of the system $|gs\rangle\simeq U(\vec\theta^*) \ket{\phi_0}$ with the ground state energy being $E_{gs}\simeq E(\vec\theta^*)$.

\section{Robustness analysis}

One of the key issues in quantum control is the robustness of the algorithm against noise. There are two notions of robustness which need to be considered in realization of quantum algorithms. One which is called imperfection-robustness focuses on the performance of the algorithm in the presence of noise. The sources of this noise include: (i) imperfections in applying control fields, such as random uncertainty in $\vec\theta$ in the VQE algorithm; or (ii) the presence of dephasing in the quantum simulator. The outcome of an imperfection-robust algorithm is expected to remain reasonably well in the presence of either of these noises.

There is, however, another notion of robustness which deals with training. In this situation, someone may train a set of parameters for a quantum control algorithm, e.g. $\vec\theta$ in the VQE, on a special quantum simulator. An important question is whether such parameters can still be used on a different quantum simulator with different noise strengths. We call this training-robustness. Note that in practice, quantum simulators are now available in various physical platforms which have different types and strengths of noise. Therefore, it is highly important to see whether a quantum control algorithm, trained on a quantum simulator, can still be used on another simulator with different strength (and even type) of noise. For the VQE algorithm, the training procedure is very time-consuming. Therefore, it is of most importance to see whether a trained circuit can directly be used on other quantum simulators. 

A quantum algorithm such as the VQE should ideally provide both imperfection- and training-robustness together. In particular, it is important to see whether the conventional circuit designs, such as the HEA, can provide such robustness in practice.

\section{Model}

In this paper, we focus on two important models which have been extensively used in the context of VQE, namely the hydrogen molecule~\cite{cao2019quantum} and the Heisenberg Hamiltonian~\cite{endoh1974dynamics}. In order to obtain a qubit Hamiltonian for the hydrogen molecule, one needs to apply Jordan-Wigner transformation to the fermionic model. The result can be described by a four-qubit Hamiltonian
\begin{align}\label{chemicaleq}
	H_{\text{h}}=&g_0+\sum_{i=0}^3g_i \sigma_{z}^{i}+\sum_{i=1,k=1,i<k}^3 g_{i,k} \sigma_{z}^{i} \sigma_{z}^{k} \nonumber\\
	&+ g_{a}\sigma_{y}^{0} \sigma_{x}^{1} \sigma_{x}^{2} \sigma_{y}^{3} + g_{b} \sigma_{y}^{0} \sigma_{y}^{1} \sigma_{x}^{2} \sigma_{x}^{3}\nonumber\\
	&+ g_{c} \sigma_{x}^{0} \sigma_{x}^{1} \sigma_{y}^{2} \sigma_{y}^{3}+ g_{d} \sigma_{x}^{0} \sigma_{y}^{1} \sigma_{y}^{2} \sigma_{x}^{3},
\end{align}
where $\sigma_\alpha^i$ (for $\alpha=x,y,z$) denotes the Pauli operator $\alpha$ acting on the $i$-th qubit and $\{g_i,g_{i,k}\}$ are the couplings determined by the hydrogen-hydrogen bond length~\cite{du2020quantum}.

To show the generality of our findings, apart from the hydrogen molecule Hamiltonian in Eq.~(\ref{chemicaleq}), we also consider the anti-ferromagnetic Heisenberg Hamiltonian 
\begin{equation}\label{Heisenbergeq}
	H_{\text{AF}} = J\sum_{i=1}^{N - 1} \left(\sigma_x^{i} \sigma_x^{i + 1}+\sigma_y^{i}  \sigma_y^{i + 1}+\sigma_z^{i} \sigma_z^{i + 1} \right),
\end{equation}
where $J>0$ is the exchange coupling. The Heisenberg Hamiltonian lies at the foundation of condensed matter physics and has been studied as a standard model for VQE algorithms~\cite{peruzzo2014variational,lyu2020accelerated,jattana2022improved}.

\section{Robustness of HEA}

As mentioned before, a truly robust quantum variational algorithm, including the VQE, must to be resilient against both imperfections and training. In the imperfection-robustness scheme, one sees the performance of the algorithm on a noisy quantum simulator and verifies its robustness against noise in such a device. In this scheme, one expects that the outcomes change very slowly as the noise increases in the device. In the training-robustness scheme, however, one needs to investigate whether a trained circuit on one quantum simulator can still perform well on another device with different noise strengths. A truly robust algorithm must to perform well in both of these two notions of robustness. 

In practice, there is always uncertainty in the precision of pulses that are used to generate in a quantum circuit to perform operations. This means that in each iteration of the VQE algorithm the parameters $\vec\theta^{(k)}$ cannot be applied very precisely and instead what is performed in the circuit is $\vec\theta^{(k)}+\delta_{\vec\theta}^{(k)}$, where $\delta_{\vec\theta}^{(k)}=(\delta_{\theta_1}^{(k)},\delta_{\theta_2}^{(k)},\cdots,\delta_{\theta_M}^{(k)})$ is a vector of random variables sampled from a normal distribution with average zero and variance $\Delta$, namely $\delta_{\theta_i}^{(k)} \sim \mathcal{N}(0,\Delta)$. The uncertainty in $\vec\theta^{(k)}$ does not change the unitarity of the evolution and thus we call it \emph{coherent noise}. Note that, despite preserving the unitarity of the evolution, this coherent noise still creates uncertainty at the output of the quantum circuit and thus affects the quality of the VQE. The coherent noise changes the operation of the quantum circuit to $U(\vec\theta^{(k)}+\delta_{\vec\theta}^{(k)})\ket{\phi_0}$. 
This means that the average energy which is truly measured in a noisy quantum simulator at the $k$-th iteration is
\begin{equation}\label{eq:coherent_cost_VQE}
	E_{\text{coh}}(\Delta|\vec\theta^{(k)})=\overline{\langle \phi_0|U^\dagger(\vec\theta^{(k)}+\delta_\vec\theta^{(k)})HU(\vec\theta^{(k)}+\delta_\vec\theta^{(k)})|\phi_0\rangle} 
\end{equation} 
where $\overline{\bullet}$ represents averaging with respect to random variable $\delta_{\vec\theta}^{(k)}$, whose elements are sampled from $\mathcal{N}(0,\Delta)$. During the training, we compute this averaging through $50$ different random samples. 
Thus, the fidelity between the output of the quantum simulator and the real ground state at each iteration $k$ becomes
\begin{equation} \label{eq:coherent_fidelity_k}
	F_{\text{coh}}(\Delta|\vec\theta^{(k)})=\overline{|\langle gs|U(\vec\theta^{(k)}+\delta_\vec\theta^{(k)})|\phi_0\rangle|^2}.
\end{equation} 
Since the measured average energy is different from the ideal case of Eq.~(\ref{eq:cost_VQE}), the final optimal parameters will also change to $\vec\theta^*_{\Delta_{\text{t}}}$, where $\Delta_{\text{t}}$ represents the uncertainty strength during the training. Therefore, in the imperfection-robustness scheme, where the whole process takes place in a single quantum simulator, the final average energy and the fidelity becomes
\begin{eqnarray}
	E_{\text{IR}}^*(\Delta_\text{t})&=&E_{\text{coh}}(\Delta_\text{t}|\vec\theta^*_{\Delta_{\text{t}}}) \cr
	F_{\text{IR}}^*(\Delta_\text{t})&=&	F_{\text{coh}}(\Delta_\text{t}|\vec\theta^*_{\Delta_\text{t}}).
\end{eqnarray}
Note that to compute $E_{\text{IR}}^*(\Delta_\text{t})$ and $F_{\text{IR}}^*(\Delta_\text{t})$ precisely, once the VQE circuit is trained, we average the output over $\sim 10^4$ different random samples, see Eqs.(\ref{eq:coherent_cost_VQE}) and (\ref{eq:coherent_fidelity_k}).

In the case of training-robustness, however, the training procedure is accomplished in a quantum simulator which results in optimal parameters $\vec\theta^*_{\Delta_{\text{t}}}$. However, this circuit might be used in a different quantum simulator in which the uncertainty strength $\Delta$ is not necessarily equal to $\Delta_\text{t}$. In this case, the average energy and the obtainable fidelity for training-robustness are given by
\begin{eqnarray}
	E_{\text{TR}}^*(\Delta|\Delta_\text{t})&=& E_{\text{coh}}(\Delta|\vec\theta^*_{\Delta_\text{t}}) \cr
	F_{\text{TR}}^*(\Delta|\Delta_\text{t})&=&	F_{\text{coh}}(\Delta|\vec\theta^*_{\Delta_\text{t}}).
\end{eqnarray} 
Note that in $E_{\text{coh}}(\Delta|\vec\theta^*_{\Delta_\text{t}})$ and $F_{\text{coh}}(\Delta|\vec\theta^*_{\Delta_\text{t}})$, the parameter $\Delta$ determines the uncertainty strength in the quantum simulator which uses the optimized VQE circuit, while $\Delta_\text{t}$ represents the uncertainty strength in the training quantum simulator. Similar to the imperfection-robustness, to compute $E_{\text{TR}}^*(\Delta|\Delta_\text{t})$ and $F_{\text{TR}}^*(\Delta|\Delta_\text{t})$ the averaging process is accomplished over a trained VQE circuit by selecting $\sim 10^4$ random samples, see Eqs.(\ref{eq:coherent_cost_VQE}) and (\ref{eq:coherent_fidelity_k}).

\begin{figure}[htp]
	\centering
	\includegraphics[width=0.7\columnwidth]{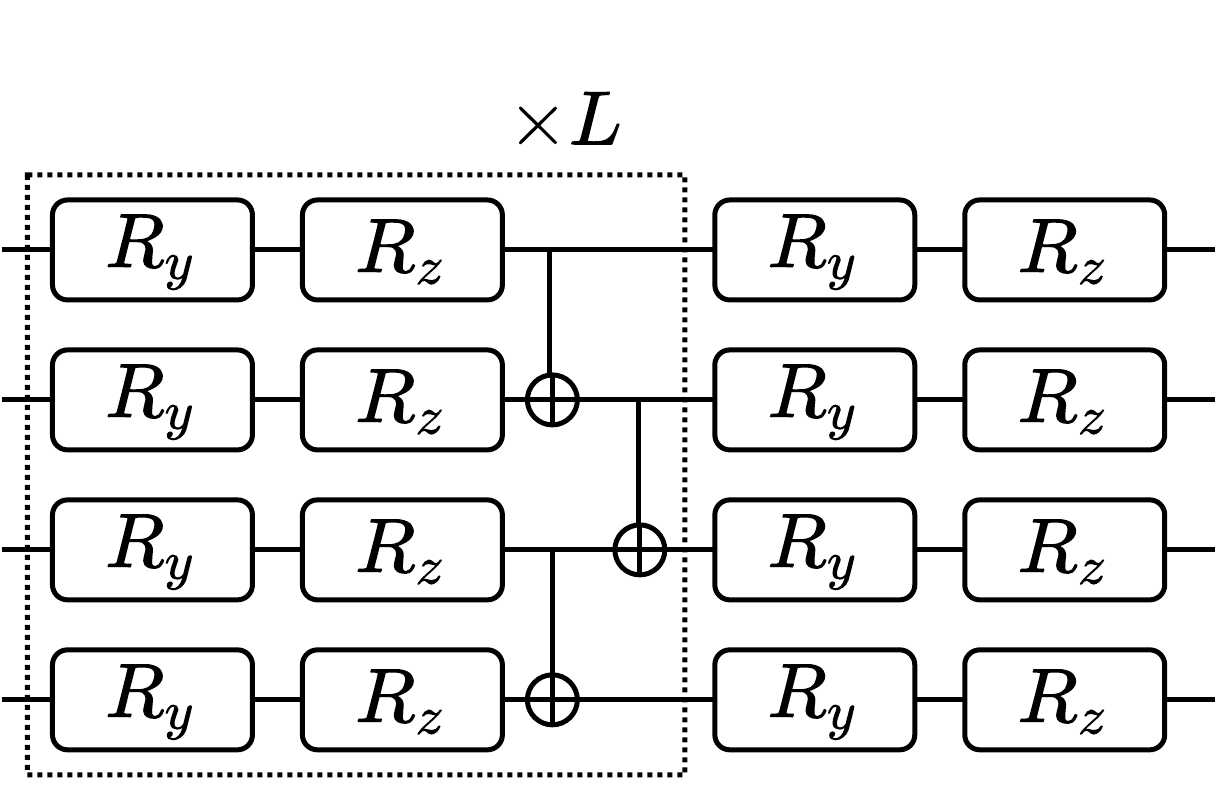}
	\caption{The hardware efficient ansatz for a $4$-qubit system with $L$ layers where $R_y=e^{\frac{-i\theta\sigma_y}{2}}$ and $R_z=e^{\frac{-i\theta\sigma_z}{2}}$ represent the parameterized single rotation gates around the $y$-axis and $z$-axis, respectively. The two-qubit entangling gates are CX.}
	\label{fig:HEA_circuit}
\end{figure}

\begin{figure}[htbp]
    \centering
    \subfigcapskip=-109pt
    \subfigure[]{}{
    \begin{minipage}[t]{0.22\textwidth}
    \centering
    \includegraphics[width=1.16\textwidth]{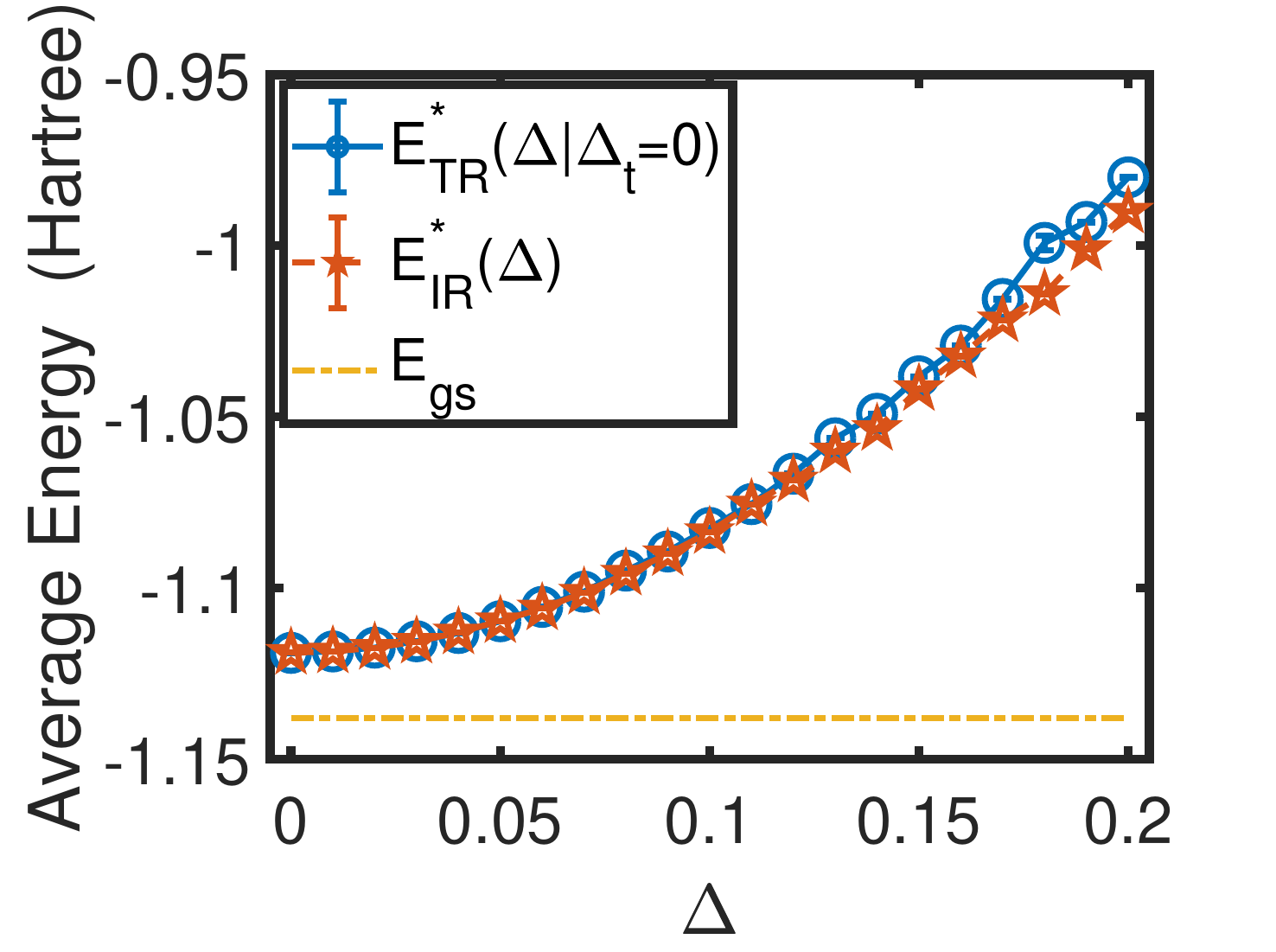}
    \end{minipage}\label{fig:HEA_Delta_a}}
    \hspace{2mm}
    \subfigure[]{}{
    \begin{minipage}[t]{0.22\textwidth}
    \centering
    \includegraphics[width=1.16\textwidth]{Fig2_HEA_Delta_b.pdf}
    \end{minipage}\label{fig:HEA_Delta_b}}
    \\
     \subfigure[]{}{
    \begin{minipage}[t]{0.22\textwidth}
    \centering
    \includegraphics[width=1.16\textwidth]{Fig2_HEA_Delta_c.pdf}
    \end{minipage}\label{fig:HEA_Delta_c}}
    \hspace{2mm}
    \subfigure[]{}{
    \begin{minipage}[t]{0.22\textwidth}
    \centering
    \includegraphics[width=1.16\textwidth]{Fig2_HEA_Delta_d.pdf}
 
    \end{minipage}\label{fig:HEA_Delta_d}}
    \caption{The performances of the HEA in approximating the ground energies of the hydrogen molecule (a)-(b) and the Heisenberg model (c)-(d) as the coherent noise $\Delta$ increases. Note that Hartree$=627.51$kcal/mol is a unit of molecular energy and $J$ is the strength of exchange coupling in the Heisenberg Hamiltonian.}
   \label{fig:HEA_Delta}
\end{figure}

HEA is one of the well-known circuit ansatzes used in VQE, as shown in Fig.~\ref{fig:HEA_circuit}. Note that the structure of the HEA is independent of the Hamiltonian and only the number of layers can vary for different Hamiltonians. 
We investigate both imperfection- and training-robustness for the HEA ansatz. In Figs.~\ref{fig:HEA_Delta}(a)-(b) we plot the average energy and the obtainable fidelity as a function of uncertainty strength $\Delta$ for the hydrogen molecule Hamiltonian. The HEA needs two layers to converge for the hydrogen molecule. In Figs.~\ref{fig:HEA_Delta}(c)-(d) the same quantities are plotted for the Heisenberg Hamiltonian which needs three layers for convergence to a high fidelity. Note that, in the case of imperfection-robustness at every single point a new set of optimal parameters are used in our HEA. As the figures show the average energy $E_{\text{IR}}^*(\Delta_{\text{t}})$ (and the fidelity $F_{\text{IR}}^*(\Delta_{\text{t}})$) increases (decreases) very rapidly as $\Delta_{\text{t}}$ increases. This shows that the HEA has little imperfection-robustness. As expected, the HEA does not show good training-robustness either. The average energy $E_{\text{TR}}^*(\Delta|\Delta_{\text{t}}=0)$ and equivalently the fidelity $F_{\text{TR}}^*(\Delta|\Delta_{\text{t}}=0)$, for which the training simulator is assumed to be perfect (i.e. $\Delta_{\text{t}}=0$), significantly vary as $\Delta$ increases indicating that the HEA does not provide training-robustness.

\begin{figure}[htbp]
    \centering
    \subfigcapskip=-109pt
    \subfigure[]{}{
    \begin{minipage}[t]{0.22\textwidth}
    \centering
    \includegraphics[width=1.16\textwidth]{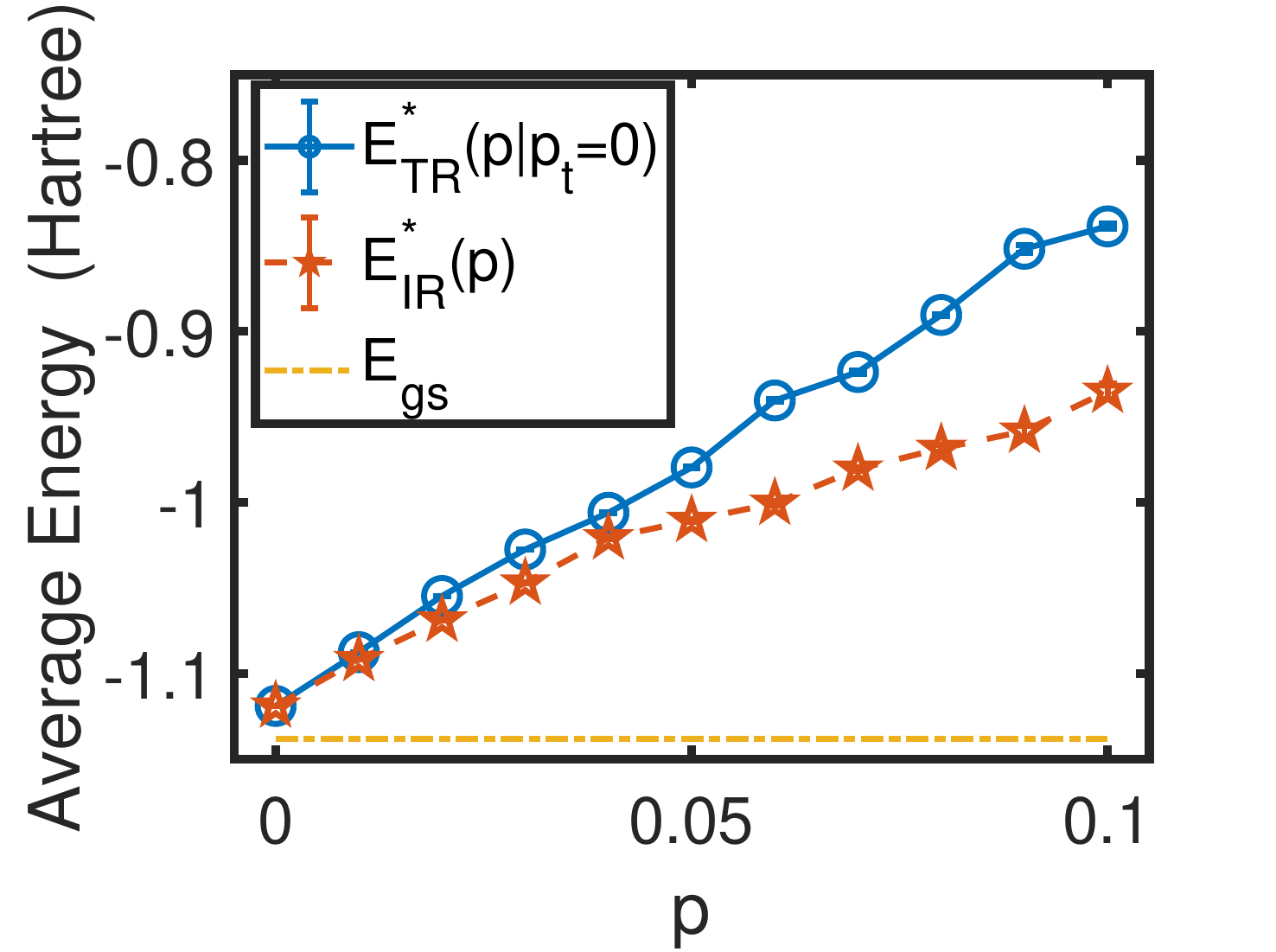}
    \end{minipage}\label{fig:HEA_p_a}}
    \hspace{2mm}
    \subfigure[]{}{
    \begin{minipage}[t]{0.22\textwidth}
    \centering
    \includegraphics[width=1.16\textwidth]{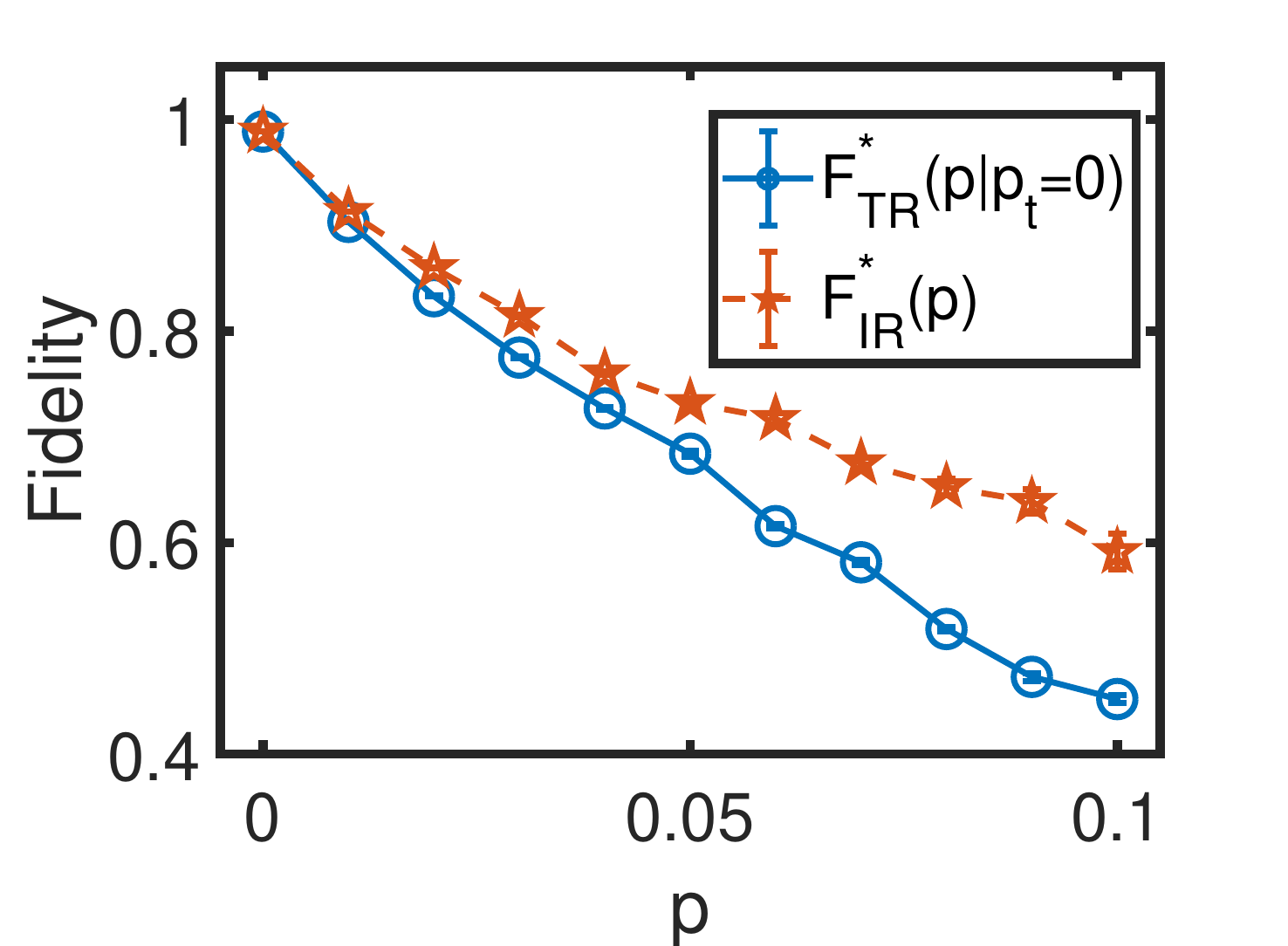}
    \end{minipage}\label{fig:HEA_p_b}}
    \\
     \subfigure[]{}{
    \begin{minipage}[t]{0.22\textwidth}
    \centering
    \includegraphics[width=1.16\textwidth]{Fig3_HEA_p_c.pdf}
    \end{minipage}\label{fig:HEA_p_c}}
    \hspace{2mm}
    \subfigure[]{}{
    \begin{minipage}[t]{0.22\textwidth}
    \centering
    \includegraphics[width=1.16\textwidth]{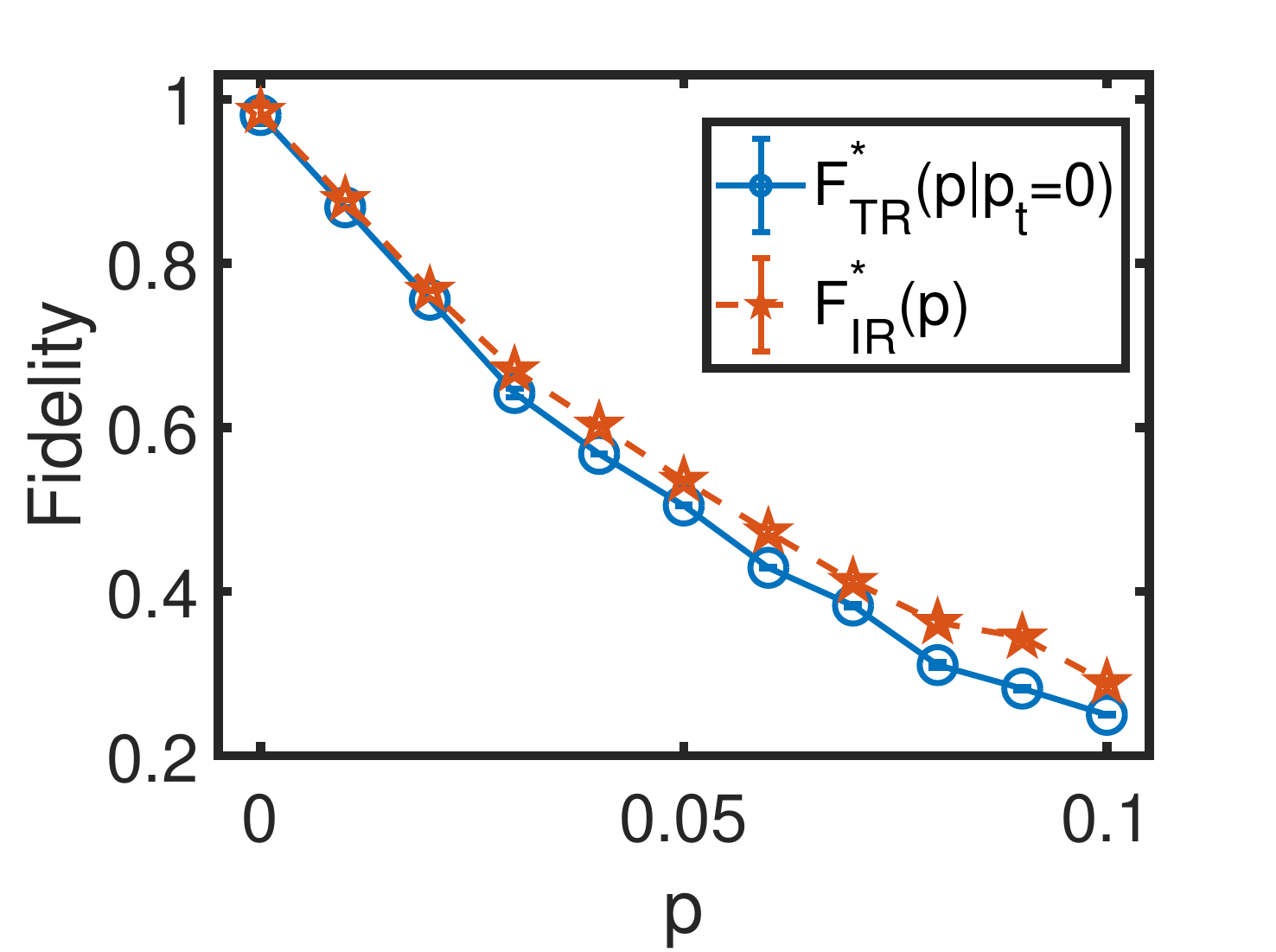}
 
    \end{minipage}\label{fig:HEA_p_d}}
    \caption{The robust performances of the HEA in approximating the ground energies of the hydrogen molecule (a)-(b) and the Heisenberg model (c)-(d) as the dephasing rate $p$ increases.}
   \label{fig:HEA_p}
\end{figure}

The coherent noise is not the only source of imperfections in quantum simulators. In fact, in NISQ devices, the incoherent noisy gate operations are the most serious source of error. Most of dephasing in NISQ simulators is resulted from the operation of the two-qubit gates. In this paper, we use Controlled-X (CX) gate as our two-qubit entangling gate whose operation is determined as $U_\text{cx}=|0\rangle \langle 0| \otimes I + |1\rangle \langle 1| \otimes \sigma_x$, where $I$ is the $2\times 2$ identity matrix. The noisy CX gate is thus described by the following operation
\begin{equation}\label{eq:incoherent_noise}
	\xi_{cx}(\rho)=(1-p)U_\text{cx} \rho U_\text{cx}^\dagger+p\sum_{i,j}| ij \rangle \langle ij | \rho | ij \rangle \langle ij |,
\end{equation}
where $p$ is the dephasing rate. For such noisy gate operations the output of the circuit is no longer a pure state and thus is determined by the density matrix $\rho_p(\vec\theta^{(k)})$. By performing the VQE on a noisy quantum simulator with incoherent gates of Eq.~(\ref{eq:incoherent_noise}) the measured average energy and the fidelity at iteration $k$ becomes 
\begin{eqnarray}\label{eq:cost_dephasing}
	E_{\text{inc}}(p|\vec\theta^{(k)})&=&\Tr\left[{H\rho_p(\vec\theta^{(k)})} \right] \cr
	F_{\text{inc}}(p|\vec\theta^{(k)})&=&\langle gs|\rho_p(\vec\theta^{(k)})|gs\rangle.
\end{eqnarray}
We assume that the quantum simulator on which the training is performed has dephasing rate of $p_{\text{t}}$. The minimization of the average energy with respect to $\vec\theta^{(k)}$ as the cost function of the VQE leads to the optimal parameter $\vec\theta^*_{p_{\text{t}}}$. For the case of imperfection-robustness, after the training one can investigate the final average energy and the fidelity
\begin{eqnarray}
	E_{\text{IR}}^*(p_{\text{t}})&=&E_{\text{inc}}(p_{\text{t}}|\vec\theta^*_{p_{\text{t}}}) \cr
	F_{\text{IR}}^*(p_{\text{t}})&=&F_{\text{inc}}(p_{\text{t}}|\vec\theta^*_{p_{\text{t}}}).
\end{eqnarray}
In the case of training-robustness, the optimized VQE circuit which has been trained on a quantum simulator with $p_{\text{t}}$ is used in another simulator with dephasing rate $p$. In this case, the average energy and the fidelity becomes
 \begin{eqnarray}
 	E_{\text{TR}}^*(p|p_{\text{t}})&=&E_{\text{inc}}(p|\vec\theta^*_{p_{\text{t}}}) \cr
 	F_{\text{TR}}^*(p|p_{\text{t}})&=&F_{\text{inc}}(p|\vec\theta^*_{p_{\text{t}}}).
 \end{eqnarray}
In Figs.~\ref{fig:HEA_p}(a)-(b), for the hydrogen molecule, we plot the average energy and the fidelity as a function of dephasing rate $p$, respectively. The similar plots for the Heisenberg Hamiltonian are plotted in Figs.~\ref{fig:HEA_p}(c)-(d). The average energy $E_{\text{IR}}^*(p_{\text{t}})$ and its corresponding fidelity $F_{\text{IR}}^*(p_{\text{t}})$ change significantly as $p_{\text{t}}$ varies showing that the HEA does not have an imperfection-robustness against dephasing. For the case of training-robustness, we assume that the training is performed on an ideal simulator with $p_{\text{t}}=0$ and the circuit is used on another quantum simulator with dephasing rate $p$. The performances of $E_{\text{TR}}^*(p|p_\text{t}=0)$ and $F_{\text{TR}}^*(p|p_\text{t}=0)$ as a function of $p$ are plotted in Fig.~\ref{fig:HEA_p}. Clearly, the performance is even worse than the case of imperfection-robustness. This clearly shows that the HEA is not a robust circuit against dephasing.

Indeed, the analysis of this section clearly shows that the HEA does not show robust behavior against both coherent and incoherent noises in any of the robustness notions. Therefore, in order to have a robust VQE performance, one needs to find a different ansatz for the quantum circuit.

\section{Robust Design: Quantum circuit evolution of augmenting topologies}

As discussed in the previous section, the widely used HEA does not provide a robust circuit against noise. In this section, we focus on the hardware to design a quantum circuit which naturally provides more robustness against noise. In particular, we develop an evolutionary algorithm which finds the best circuit design, inspired by natural selection. Finding the optimal circuit is a non-smooth and non-convex optimization process which makes it very challenging. It is known that for such non-smooth optimization problems the evolutionary algorithms provide useful heuristic solutions~\cite{coello2007evolutionary}. Hence, we aim to formulate an evolutionary algorithm to optimize over different circuit ansatzes. We introduce Quantum Circuit Evolution of Augmenting Topologies (QCEAT) as an evolutionary algorithm which optimizes the quantum ansatz structure to design more robust quantum circuits. The QCEAT algorithm is inspired by Neural Evolution of Augmenting Topologies (NEAT) which is used in classical neural networks~\cite{stanley2002evolving}.

\begin{figure}[htp]
	\centering
	\includegraphics[width=1\columnwidth]{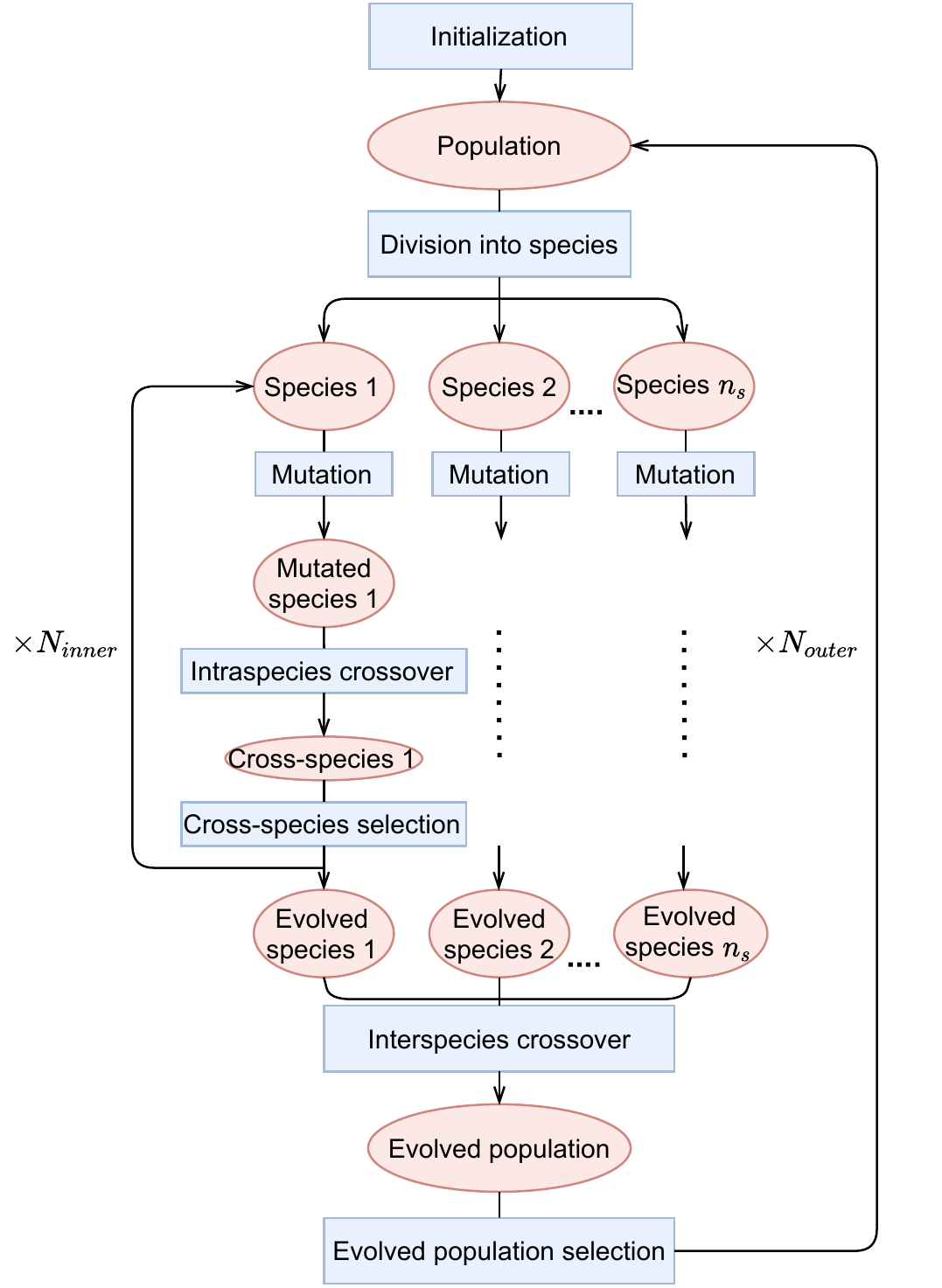}
	\caption{The flowchart of the QCEAT algorithm to generate a robust quantum circuit. The ellipses represent sets of circuits and rectangles denote operations which are performed on the circuits. $N_{\text{inner}}$ indicates the number of repetition of the inner loop and $N_{\text{outer}}$ indicates the number of repetition of the outer loop. Both of them are chosen at the beginning of the QCEAT.}
	\label{fig:QCEAT_diagram}
\end{figure}

\subsection{QCEAT algorithm}

\begin{figure}[ht]
    \centering
    \includegraphics[width=\columnwidth]{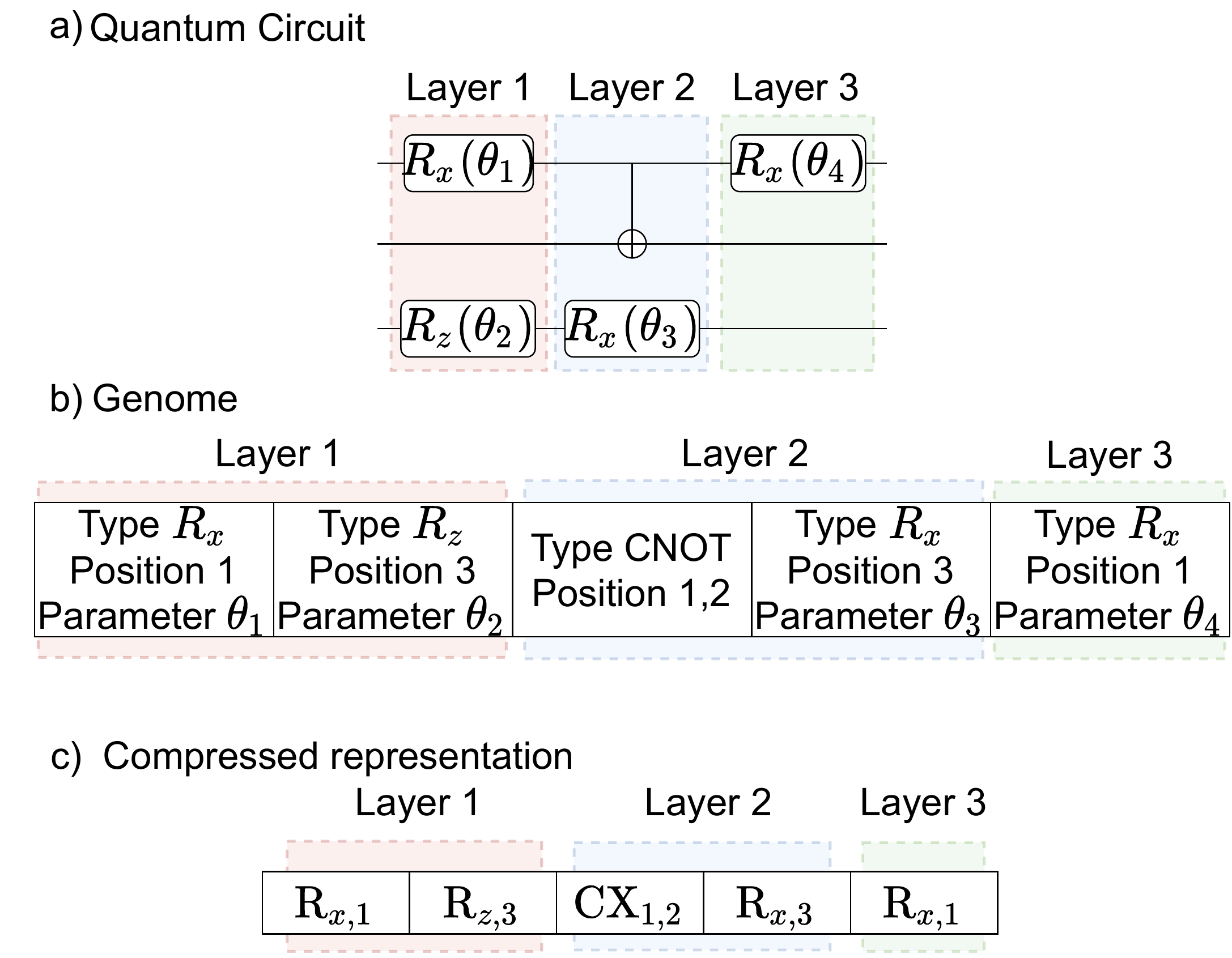}
    \caption{The encoding scheme mapping a quantum circuit to a genome. (a) An example of a $3$-layer variational quantum circuit where dashed boxes represents each layer; (b) The $3$-layer genome corresponding to the $3$-layer variational circuit. The genome elements of one layer are ordered from first qubit towards the last qubit. (c) For the sake of brevity, the presentation of genomes can be compressed which will be used in the rest of the paper.}
    \label{fig:Genome}
\end{figure}

The essence of our QCEAT algorithm is to encode each quantum circuit into a genome, and describe the variations of the circuit as mutations and crossovers within the set of encoded genomes. Then optimizing the variations of the quantum circuit structure, is equivalent to a population evolution in terms of mutations and crossovers among the set of encoded genomes, from generation to generation. The QCEAT algorithm is schematically depicted in Fig.~\ref{fig:QCEAT_diagram}, with the encoding scheme in Fig.~\ref{fig:Genome}, and the evolutionary mutations and crossovers described in Fig.~\ref{fig:Mutation_CrossOver}. Specifically, in Fig.~\ref{fig:QCEAT_diagram}, every ellipse represents a set of circuits and every rectangle describes an operation which acts on the set of circuits. In the following, we will describe the variational quantum circuits and its variations in terms of terminologies in genomics. Specifically, a \emph{population} of quantum circuits is defined as a set of variational circuits, encoded as circuit genomes; a \emph{species} of the population is defined as a subset of the population with similar circuit structure. We assume the population can be divided into $n_s$ number of different species. Each circuit genome in the population consists of a few layers of elementary gates, abstracted as \emph{genes}. The variations of the circuits are formulated as evolutionary variations of the population, which include mutation, intraspecies crossover, and interspecies crossover. After the evolutionary variations, a population of circuits will experience the natural selection process that will eliminate the worse genomes according to some fitness function evaluation. For our problem, the fitness function of a circuit is defined as the average energy of the output state of the circuit genome. We consider two types of natural selection process in this work: the cross-species selection and the evolved-population selection, with detailed explained in the next subsection. With all these definitions, we are now ready to give an overall description of the QCEAT algorithm. In Step 1, for an $N$-qubit system, a population of $N$ circuits is chosen randomly, each with $N$ different elementary gates (either single- or two-qubit ones). In this work, we use three different types of elementary gates, i.e., the single-qubit rotations $R_x(\theta)=e^{-i\sigma_x \theta/2}$ and $R_z(\theta)=e^{-i\sigma_z \theta/2}$, and the two-qubit gate CX. Note that the parameters of the single-qubit rotations are variables which are used for training the circuit. In Step 2, the population is divided into $k$ different species based on a similarity function, so that the circuits in each species share similar structure. In Step 3, each species will go through an evolutionary process, in which every circuit genome in each species will experience a mutation with a certain probability $P_{\text{mut}}$, e.g., $P_\text{mut}{=}0.7$. Such mutation can be either an addition, a deletion or a substitution of an elementary gate in the circuit, corresponding to an addition, a deletion, or a substitution of one gene in the corresponding genome. All mutated circuits will then be added to the old species set to enlarge it into a new one, denoted as the \emph{mutated species} in the diagram in Fig.~\ref{fig:QCEAT_diagram}. In Step 4, after the mutation process, every pair of circuit genomes within each mutated species will make an intraspecies crossover with probability $P_{\text{cross}}^{\text{intra}}$. All genomes generated by intraspecies crossovers within each species will then be added to the old mutated species set to enlarge it into a new one denoted as the \emph{cross-species} set. In Step 5, an optimization is performed to each circuit newly-generated in the mutation and the intra-crossover process, and then we apply the natural selection procedure and discard half of the circuits of each cross-species with the highest average energy, calculated by Eqs.~(\ref{eq:coherent_cost_VQE}) and (\ref{eq:cost_dephasing}), with $k=500$ number of optimization iterations. Up to this point, Step 3 to Step 5 form an inner loop iteration and such iteration will be repeated for $N_{\text{inner}}$ times, as shown in Fig.~\ref{fig:QCEAT_diagram}, before Step 6. After the inner loop iterations are completed, each updated species is denoted as the \emph{evolved species}.  In Step 6, we introduce the interspecies crossovers that will reproduce new circuits by probabilistically combining any pair of two circuits from two different species. Such process further increases the diversity of circuits in the population. After this step, we merge the circuits newly generated from interspecies crossovers and the circuit from all evolved species into a new population denoted as the \emph{evolved population}. In Step 7, an optimization is performed to each variational circuit newly added to the population and then a natural selection is taken again to the evolved population to discard the circuits with the highest average energy, so that the number of remaining circuits in the population is no than a threshold $n_r$, i.e., $n_r=50$. In other words, if the number of remaining circuits is large than $n_r$, we delete the circuits with the highest average energy so that the size of the population is cut to $n_r$. The process of a population passing from Step 1 to Step 7, including the $N_{\text{inner}}$ inner loop iterations from Step 3 to Step 5, is denoted as one outer loop iteration of QCEAT, and the resulting population at end of Step 7 will be used as the starting population of the next iteration. The entire QCEAT algorithm consist of $N_{\text{outer}}$ number of such outer loop iterations, as illustrated in Fig.~\ref{fig:QCEAT_diagram}. 


\subsection{Detailed operations in QCEAT}

In this subsection, we explain the operations mentioned above for describing the QCEAT algorithm in details. Those who are not interested in the details of the operations can skip this subsection. As mentioned before, the required operations in QCEAT are given in the rectangles of the diagram in Fig.~\ref{fig:QCEAT_diagram}. In the previous subsection, we explained these operations very generally without going through the details. In this part, we explain each operation in detail.

\textbf{Encoding of variational circuits into genomes:} We need to introduce an encoding that maps each variational circuit into a circuit genome, and each elementary gate into a gene. Each gene contains at most three features: (i) gate type (namely, $R_x$, $R_z$ and CX); (ii) the qubit (or qubits) on which the gate operates; and (iii) the parameter of the gate (e.g. an angle $\theta$), which is only relevant for $R_x$ and $R_z$. For any given quantum circuit, one can divide the elementary gates into layers according to their order in time. Gates in each layer can operate simultaneously. In Fig.~\ref{fig:Genome}(a), we illustrate such encoding scheme using an example of a quantum circuit on three qubits. The gates are divided into three different layers. While the quantum gates in each layer can operate simultaneously, they cannot act earlier than or at the same time as the gates in the previous layer. The information of each gene is determined by the corresponding gate of the circuit, and the order of the genes in the genome agrees with the order to the layers of gates in the circuit. Each gene contains the following information: the gate type, the qubit(s) on which the gate operates, and the gate parameter. For instance, the circuit shown in Fig.~\ref{fig:Genome}(a) is encoded into the genome in Fig.~\ref{fig:Genome}(b), and its notation can be further simplified into Fig.~\ref{fig:Genome}(c). Note that if two single-qubit gates of the same type act on the same qubit in two adjacent layers, one can absorb them into one gate in the earlier layer. In the case of two CX gates acting on the same neighboring qubits in two subsequent layers, one can remove them as their action is equal to the identity. Now we are ready to explain each operation in Fig.~\ref{fig:QCEAT_diagram}.

\textbf{Division into species:} The first operation in the QCEAT algorithm, see the diagram of Fig.~\ref{fig:QCEAT_diagram}, is to divide the population into $n_s$ species. In order to do that one needs to quantify the distance between two quantum circuits, or equivalently between their genomes. Since only the structure of the circuits is important for dividing them into species, we ignore the parameters of the gates. We define three basic actions that one can perform on a circuit genome: (i) addition; (ii) deletion; and (iii) substitution of a gate. The distance between two quantum circuits, is defined as the minimum number of basic actions that one needs to convert one of the genomes into the other. To determine this distance between any pair of genomes we use minimized edit distance algorithm~\cite{navarro2001guided} which is a string metric used to measure the difference between two sequences. 

To divide the population into different species, we assign the first genome (i.e. the quantum circuit) of the population into the first species set. Then one selects the next genome from the population. If its distance with the circuit in the first species is less than a threshold we include it into that species set. Otherwise, we assign it to a new species set. Then we continue the procedure and for every genome from the population we compute the average distance from all the members of a species set and if the average distance is below the threshold it will be added to that set. If it does not include to any of the species sets, we create a new set for that genome. To find the species set of a genome with the largest $N_{g}$ gate, we set the threshold to be $\eta \times N_{g}$, where $\eta$ is constant. We choose $\eta=0.3$ in this paper. By playing with $\eta$ one can control the number of species.

\begin{figure*}[htp]
\centering
\includegraphics[width=1\textwidth, height=0.15\textwidth]{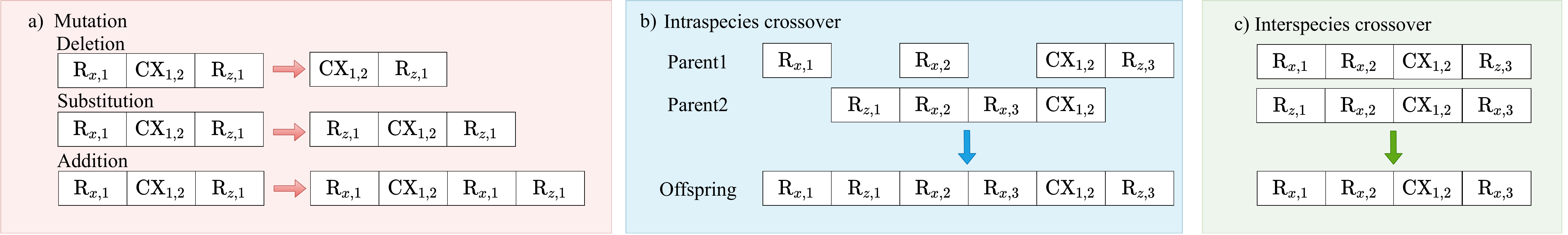}
\caption{Three basic operations for the QCEAT to change the structure of a genome (i.e., circuit): (a) mutation operations, including deletion, substitution and addition; (b) intraspecies crossover, which recombines two parent genomes from the same species to generate an offspring genome; and (c) interspecies crossover, which recombines two parent genomes from different species to generate an offspring genome.}\label{fig:Mutation_CrossOver}
\end{figure*}

\textbf{Mutation:} Every genome (i.e. circuit) in a species set may face the mutation operation with the probability $P_{\text{mut}}$, which is taken to be $P_{\text{mut}}=0.7$. If the mutation happens, it can randomly change the genome through three different actions, each with different probabilities, namely: addition (with probability $0.625$), substitution (with probability 0.25) and deletion (with probability 0.125). For addition, a new element is added to the genome. The location of the new element, the gate type and its qubit position are all randomly chosen via a uniform distribution. For substitution, one of the elements of the genome is randomly chosen and its gate type and the qubit position is randomly changed. For deletion, one element of the genome is randomly removed. The three possible mutations have been schematically explained in Fig.~\ref{fig:Mutation_CrossOver}(a).

\textbf{intraspecies crossover:} The second operation that can happen within a species set is intraspecies crossover. In this operation any pair of genomes, called \emph{parents}, from the muted species set are recombined to give birth to an \emph{offspring} genome with the probability of $P_{\text{cross}}^{\text{intra}}=0.05$. To make the intraspecies crossover, one needs to write the two parents genome in parallel to each other such that the gates of the same type and position becomes aligned, as shown in Fig.~\ref{fig:Mutation_CrossOver}(b). This can be systematically done by the longest common substring~\cite{bergroth2000survey} which is to find the longest subsequence common in two sequences. In there we use this idea to align the gates (i.e., genes). The aligned genes are called matched genes, and the others are called unmatched. The principle of intraspecies crossover is to reserve all genes (either matched or unmatched) in their order. Note that intraspecies crossover does not increase the complexity dramatically as all the genomes in a species set have similar structures with many matched genes. Thus, the offsprings produced by intraspecies crossover remain similar to the parents and in close proximity to other genomes of the species set. 

\textbf{Cross-species selection:} After mutation and intraspecies crossover, the cross-species set has a lot more genomes than the original species set. Now one needs to select the best circuits (i.e. genomes) and repeat the inner loop evolution of the species, depicted in Fig.~\ref{fig:QCEAT_diagram}. We perform the VQE algorithm for all the circuits in the cross-species sets. For coherent and incoherent noises, we use the cost functions of Eq.~(\ref{eq:coherent_cost_VQE}) and Eq.~(\ref{eq:cost_dephasing}), respectively. In each cross-species set, we keep the top $1/2$ of the circuits which have the lowest average energy and discard the remaining $1/2$ of the circuits.


\textbf{interspecies crossover:} After several repetitions of the inner loop, the evolved species set becomes very diverse with respect to the circuits. In order to enhance the diversity of circuits even further, we introduce the interspecies crossover procedure. In this operation, an interspecies crossover takes place between any pair of circuits from two different evolved species sets with the probability of $P_{\text{cross}}^{\text{inter}}=0.1$. The crossover between the genomes is different from the intraspecies crossover. In this case, the two parent genomes are aligned according to the layers. When the gates are matched the offspring inherits the same gate as well. But when the gates are not matched, one of the gates from the parents are randomly chosen to be inserted in the offspring's genome. The procedure is explained schematically in Fig.~\ref{fig:Mutation_CrossOver}(c). Since the parents are from two different species, their gate structure is very different and thus many unmatched gates exist between the parents. Therefore, the offspring genome generated from the parents can be very different. All the circuits from the species sets and the offsprings, generated by the interspecies crossover, are added to the evolved population set, as shown in Fig.~\ref{fig:QCEAT_diagram}. 

\textbf{Evolved population selection:} This selection process is similar to the cross-species selection process. Specifically, based on the value of the average energy for each circuit, we will discard the circuits with the highest average energy, so that the number of remaining circuits in the population is no than a threshold $n_r$. In our simulations, we choose $n_r=50$. Finally, after the completion of all $N_{\text{outer}}$ outer loop iterations, we select the best circuit with the lowest average energy from the final population at the end of the last outer loop iteration.

\section{Robustness of QCEAT}
In this section, we present the performance of the circuit designed by the QCEAT algorithm, as explained above, in the presence of both coherent and incoherent noise. The two notions of robustness, namely imperfection-robustness and training-robustness will be investigated for the designed circuits. The QCEAT starts with a population which contains four random circuits. At the end of the QCEAT algorithm, a total number of $n_r= 50$ circuits remain in the population among which we pick the one with the minimum average energy. In this section, we provide numerical analysis for the circuit designed by QCEAT and compare its robustness with HEA. 

\begin{figure*}[htp]
	\centering
	\includegraphics[width=2\columnwidth]{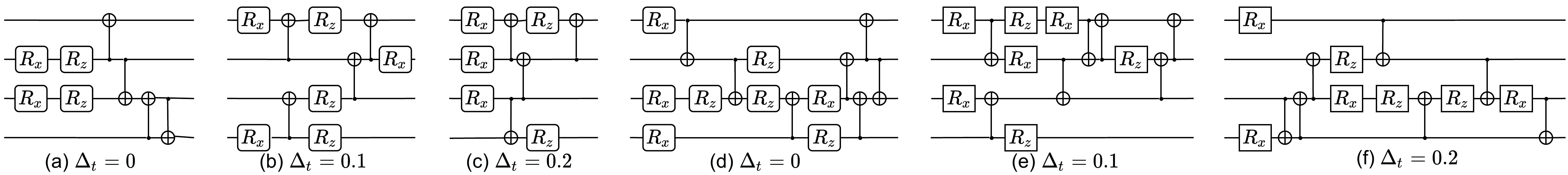}
	\caption{The optimized circuit structure generated by the QCEAT under three different values of coherent noise $\Delta_\text{t}=0,0.1,0.2$, for two models, the hydrogen molecule $H_\text{h}$ (circuits (a)-(c)), and the anti-ferromagnetic Heisenberg chain $H_{\text{AF}}$ (circuits (d)-(f)).
    }\label{fig:coherent-circuit}
\end{figure*}

\subsection{Coherent noise}

For every $\Delta$, the QCEAT provides a different design of the circuit. In Figs.~\ref{fig:coherent-circuit}(a)-(c) we present three designs of the circuit, for simulating the hydrogen molecule with $\Delta_\text{t}=0$, $\Delta_\text{t}=0.1$ and $\Delta_\text{t}=0.2$, respectively. In Figs.~\ref{fig:coherent-circuit}(d)-(f) we also depict the corresponding circuits for the Heisenberg Hamiltonian for $\Delta_\text{t}=0$, $\Delta_\text{t}=0.1$ and $\Delta_\text{t}=0.2$, respectively. Clearly, the QCEAT circuits are much simpler than the HEA and require fewer gates.

\begin{figure}[htbp]
    \centering
    \subfigcapskip=-110pt
    \subfigure[]{}{
    \begin{minipage}[t]{0.2\textwidth}
    \centering
    \includegraphics[width=1.25\textwidth]{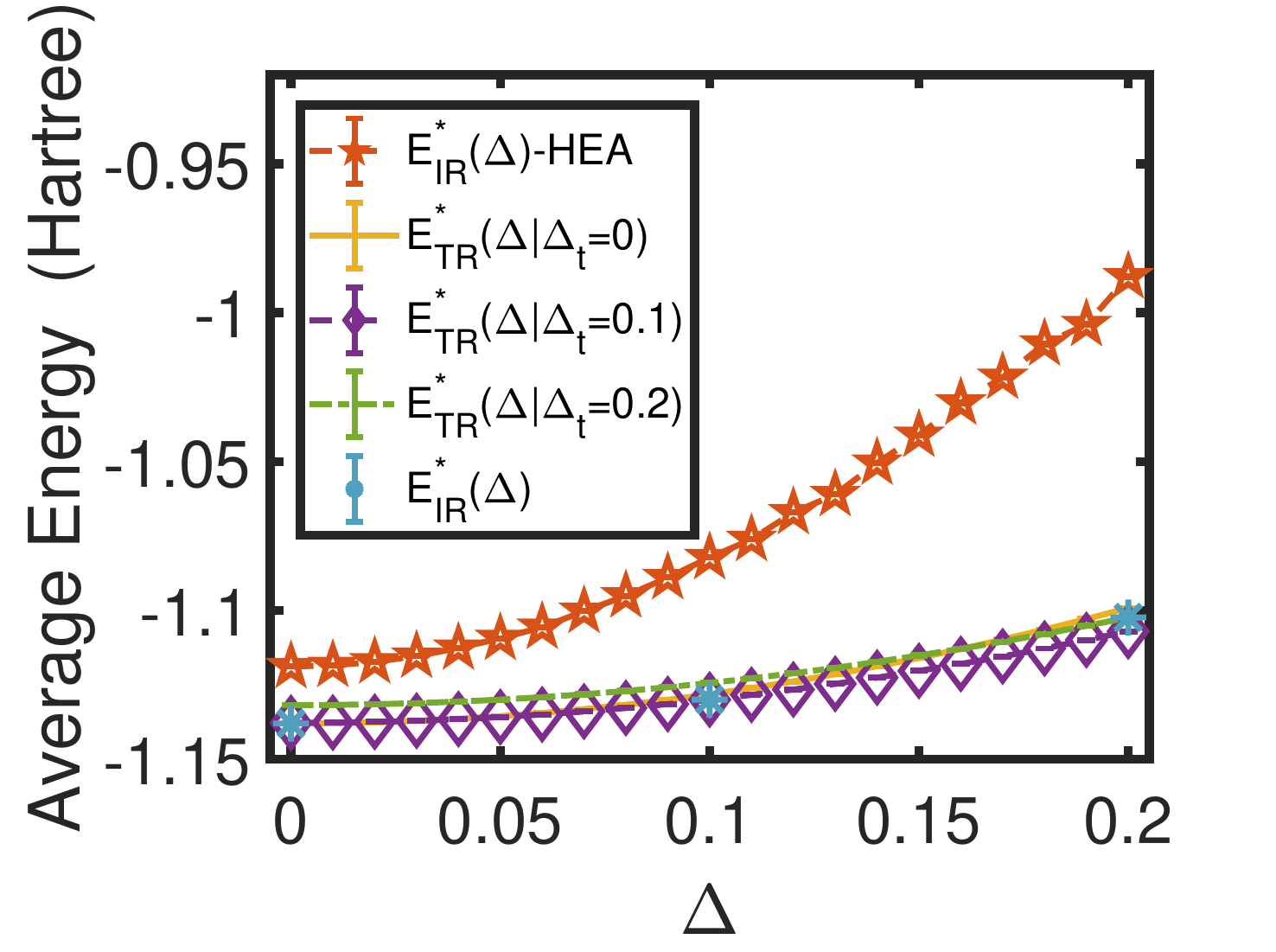}
    \end{minipage}\label{fig:Delta-QCEAT_a}}
    \hspace{0.2cm}
    \subfigure[]{}{
    \begin{minipage}[t]{0.2\textwidth}
    \centering
    \includegraphics[width=1.25\textwidth]{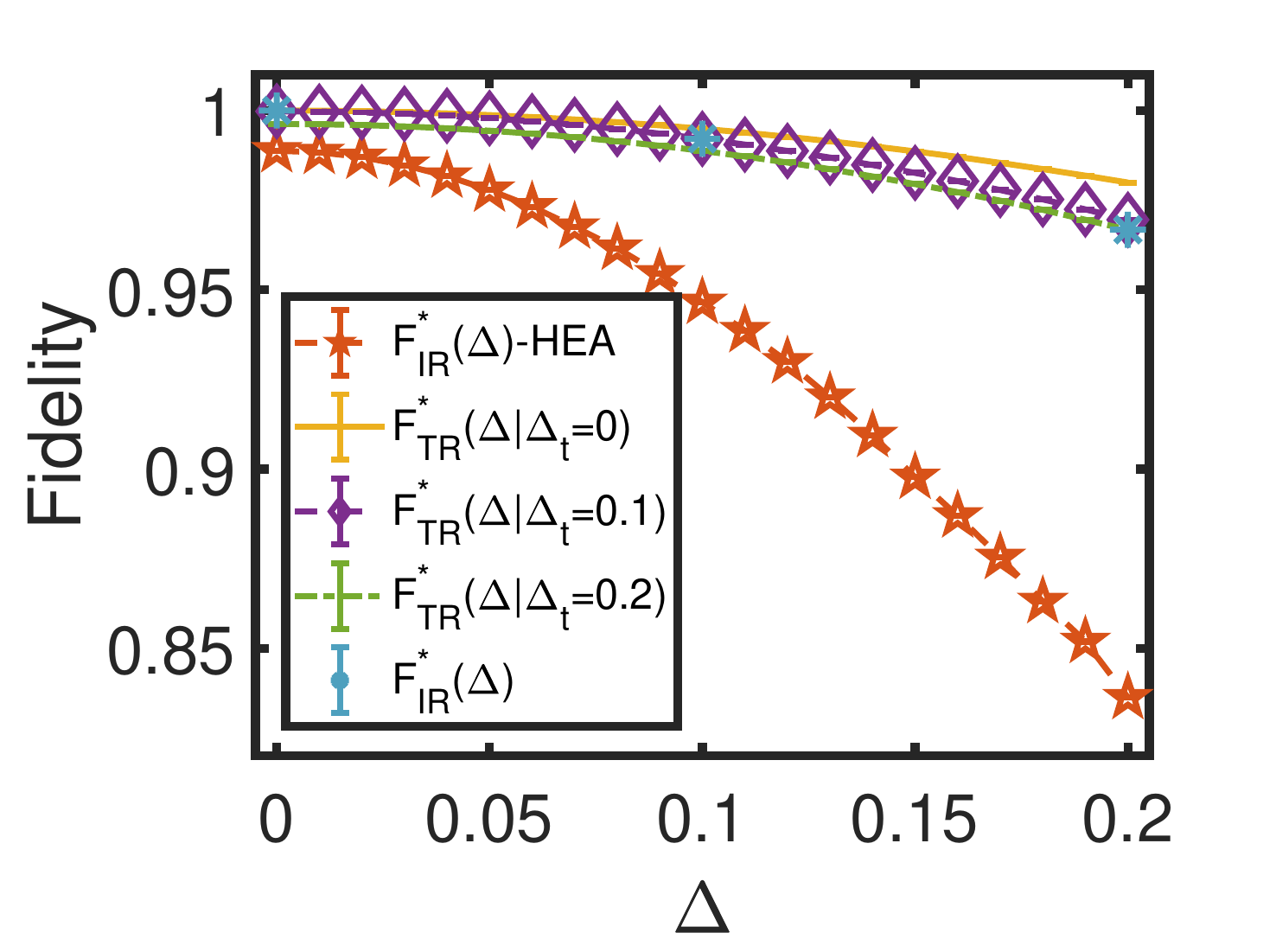}
    \end{minipage}\label{fig:Delta-QCEAT_b}}
    \\
     \subfigure[]{}{
    \begin{minipage}[t]{0.2\textwidth}
    \centering
    \includegraphics[width=1.25\textwidth]{Fig8_Delta_QCEAT_c.pdf}
    \end{minipage}\label{fig:Delta-Delta_c}}
    \hspace{0.2cm}
    \subfigure[]{}{
    \begin{minipage}[t]{0.2\textwidth}
    \centering
    \includegraphics[width=1.25\textwidth]{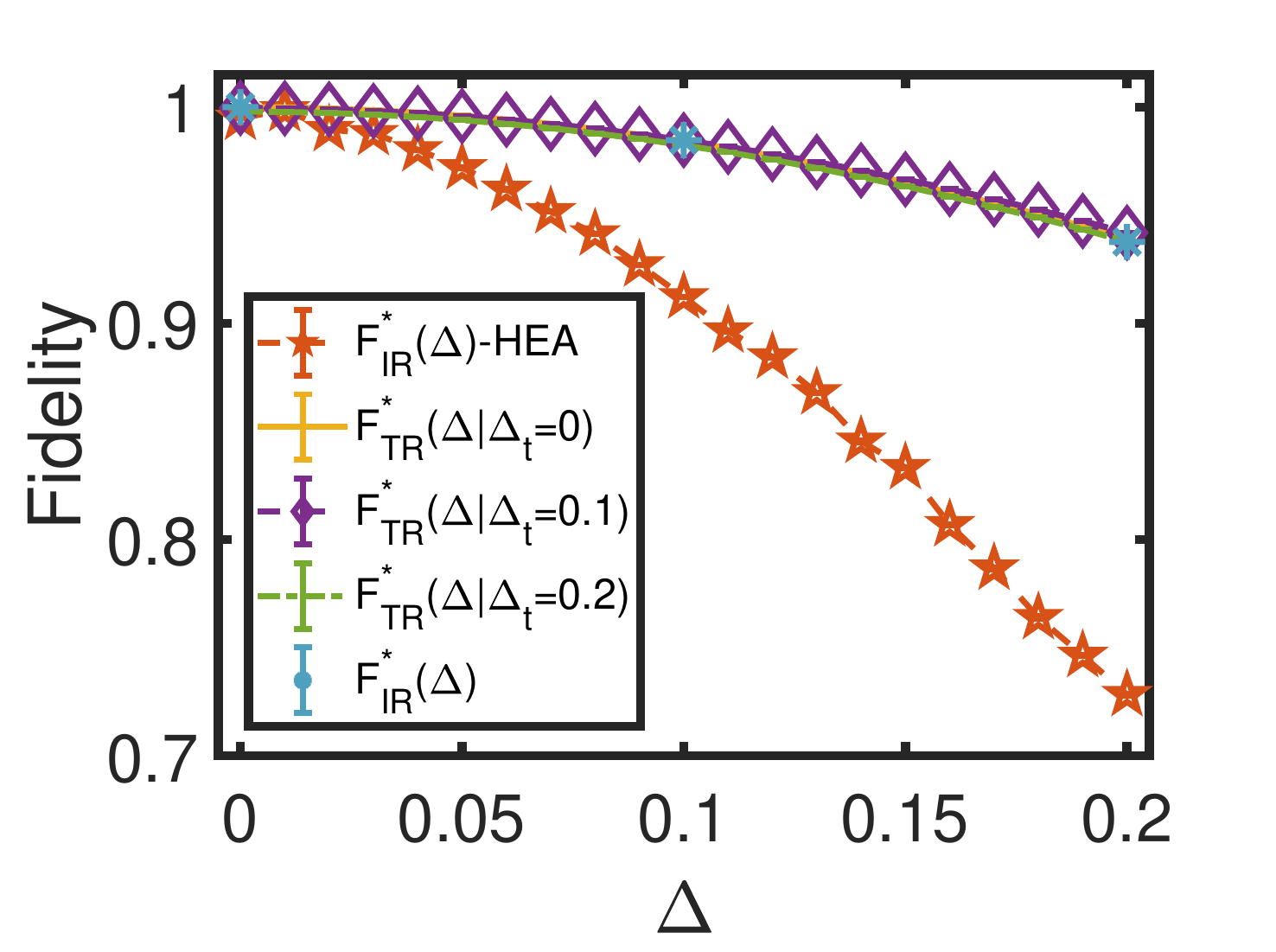}
 
    \end{minipage}\label{fig:Delta-QCEAT_d}}
    \caption{Comparison of the robustness of the HEA and the circuits generated by QCEAT (Fig.~\ref{fig:coherent-circuit}) in approximating the ground energies for the hydrogen molecule (a)-(b) and the Heisenberg model (c)-(d) as the coherent noise $\Delta$ increases.}
   \label{fig:Delta-QCEAT}
\end{figure}

To analyze the imperfection-robustness of the QCEAT circuits, in Figs.~\ref{fig:Delta-QCEAT}(a)-(b) we plot the average energy and the final fidelity as a function of $\Delta$ for the hydrogen molecule. The corresponding figures for the Heisenberg Hamiltonian are plotted in Figs.~\ref{fig:Delta-QCEAT}(c)-(d). In such analysis for every value of $\Delta$ a new circuit is trained. As the figures show the performances of QCEAT circuits are significantly more robust than the HEA, although the training has only been focused on minimization of average energy. As $\Delta$ increases, the average energy ($E_{\text{IR}}^*(\Delta_\text{t})$) and the fidelity ($F_{\text{IR}}^*(\Delta_\text{t})$) slightly change in both models across the whole range of studied $\Delta_\text{t}$. This is a clear evidence for imperfection-robustness in comparison with the HEA in which the performance significantly deteriorates by increasing $\Delta_\text{t}$.  

One can also investigate the training-robustness in which a circuit designed for a given $\Delta$ is used in other simulators with different $\Delta$. The results for average energy and the fidelity as a function of $\Delta$ are also shown in Fig.~\ref{fig:Delta-QCEAT}. Remarkably, $E_{\text{TR}}^*(\Delta|\Delta_\text{t})$ and $F_{\text{TR}}^*(\Delta|\Delta_\text{t})$ not only reveal good performance, namely low average energy and high fidelity, but also remain very close to each other indicating strong training-robustness. In other words, one can use any of the trained circuits on different quantum simulators and the outcomes perform almost equally well. 

To understand the reason behind this improvement, one needs to notice that the QCEAT algorithm naturally minimizes the number of gates in comparison with the HEA which uses two layers for the hydrogen molecule (with $24$ single-qubit gates and $6$ CX gates) and three layers for the Heisenberg Hamiltonian (with $32$ single-qubit gates and $9$ CX gates). For the coherent case, having less single-qubit gates systematically minimizes the destructive effect of such noise. In Table~\ref{table:coherent} we present the number of gates in the optimized QCEAT circuits. As the data shows the main power of the QCEAT algorithm is in minimizing the number of gates which naturally enhances the imperfection-robustness. In addition, since the circuits with various $\Delta$ are eventually optimized to have similar number of gates, their performances remain similar, indicating strong training-robustness.  

\begin{table}[htp]
    \centering
    \begin{tabular}{|c|c|c|c|c|c|c|}
    \hline
    \multirow{2}*{$\Delta_\text{t}$} &\multicolumn{2}{c|}{0} &\multicolumn{2}{c|}{0.1} & \multicolumn{2}{c|}{0.2}\\
    \cline{2-7}
     & \#R    &\#CX &\#R &\#CX &\#R & \#CX\\
    \hline
   $H_\text{h}$&4    &4 &6 &4 &5 &4\\
     \hline
    {$H_\text{AF}$}&8    &7 &7 &7 &7 &7\\
     \hline
     \end{tabular}
     \caption{The number of single- and two-qubits gates of the six circuits generated by QCEAT (Fig.~\ref{fig:coherent-circuit}) for the hydrogen molecule $H_\text{h}$ and the Heisenberg model $H_\text{AF}$ under different values of the coherent noise intensity $\Delta_\text{t}$. \#$\text{R}$ represents the number of single-qubit gates and \#$\text{CX}$ represents the number of two-qubit gates.}
     \label{table:coherent}
\end{table}

\begin{figure*}[htp]
	\centering
	\includegraphics[width=2\columnwidth]{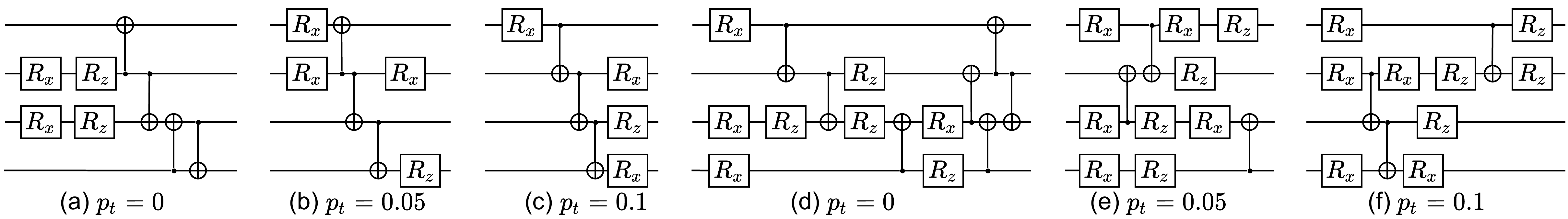}
	\caption{The optimized circuit structure generated by the QCEAT under different values of dephasing rates $p_\text{t}=0,0.05,0.1$, for two models, the hydrogen molecular $H_\text{h}$ (circuits (a)-(c)), and the Heisenberg model $H_{\text{AF}}$ (circuits (d)-(f)).
	}\label{fig:incoherent-circuit}
\end{figure*}

\subsection{Incoherent noise}

Similar to the case of coherent noise, for every value of dephasing rates $p$, the QCEAT algorithm results in different circuits. In Figs.~\ref{fig:incoherent-circuit}(a)-(c) we provide three circuits which have been obtained by the QCEAT algorithm for the hydrogen molecule with $p_\text{t}=0$, $p_\text{t}=0.05$ and $p_\text{t}=0.1$, respectively. The corresponding circuits for the Heisenberg model are given in Figs.~\ref{fig:incoherent-circuit}(d)-(f). Again what is clear from the circuits is that the number of gates is significantly smaller than the HEA which demands two layers (with $24$ single-qubit rotations and $6$ CX gates) for the hydrogen molecule and three layers (with $32$ single-qubit rotations and $9$ CX gates) for the Heisenberg model. In particular, the reduction in the number of CX gates is important as we have included all the incoherent noise in the performance of CX gates.

\begin{figure}[htbp]
    \centering
    \subfigcapskip=-110pt
    \subfigure[]{}{
    \begin{minipage}[t]{0.2\textwidth}
    \centering
    \includegraphics[width=1.25\textwidth]{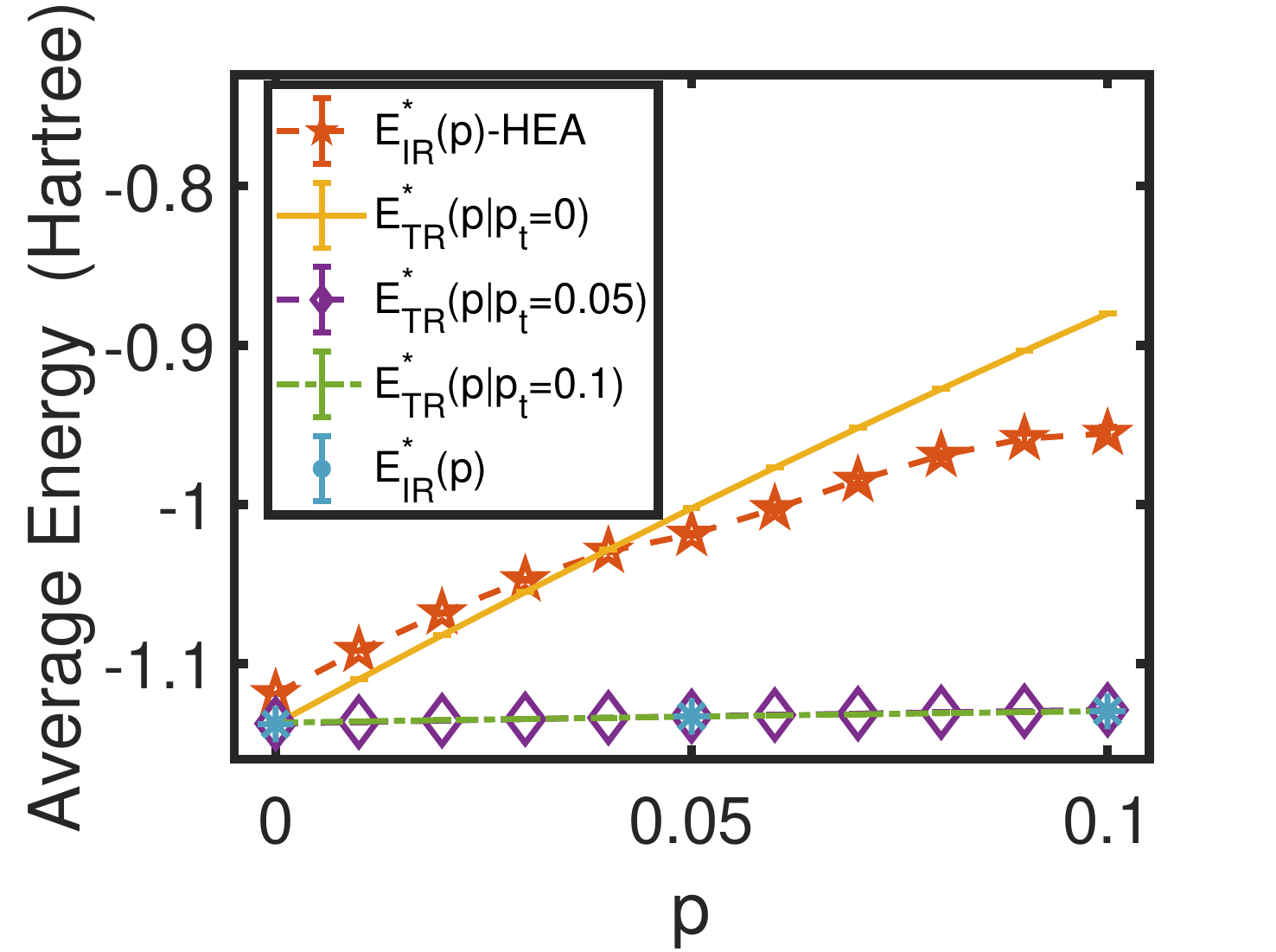}
    \end{minipage}\label{fig:p-QCEAT_a}}
    \hspace{0.2cm}
    \subfigure[]{}{
    \begin{minipage}[t]{0.2\textwidth}
    \centering
    \includegraphics[width=1.25\textwidth]{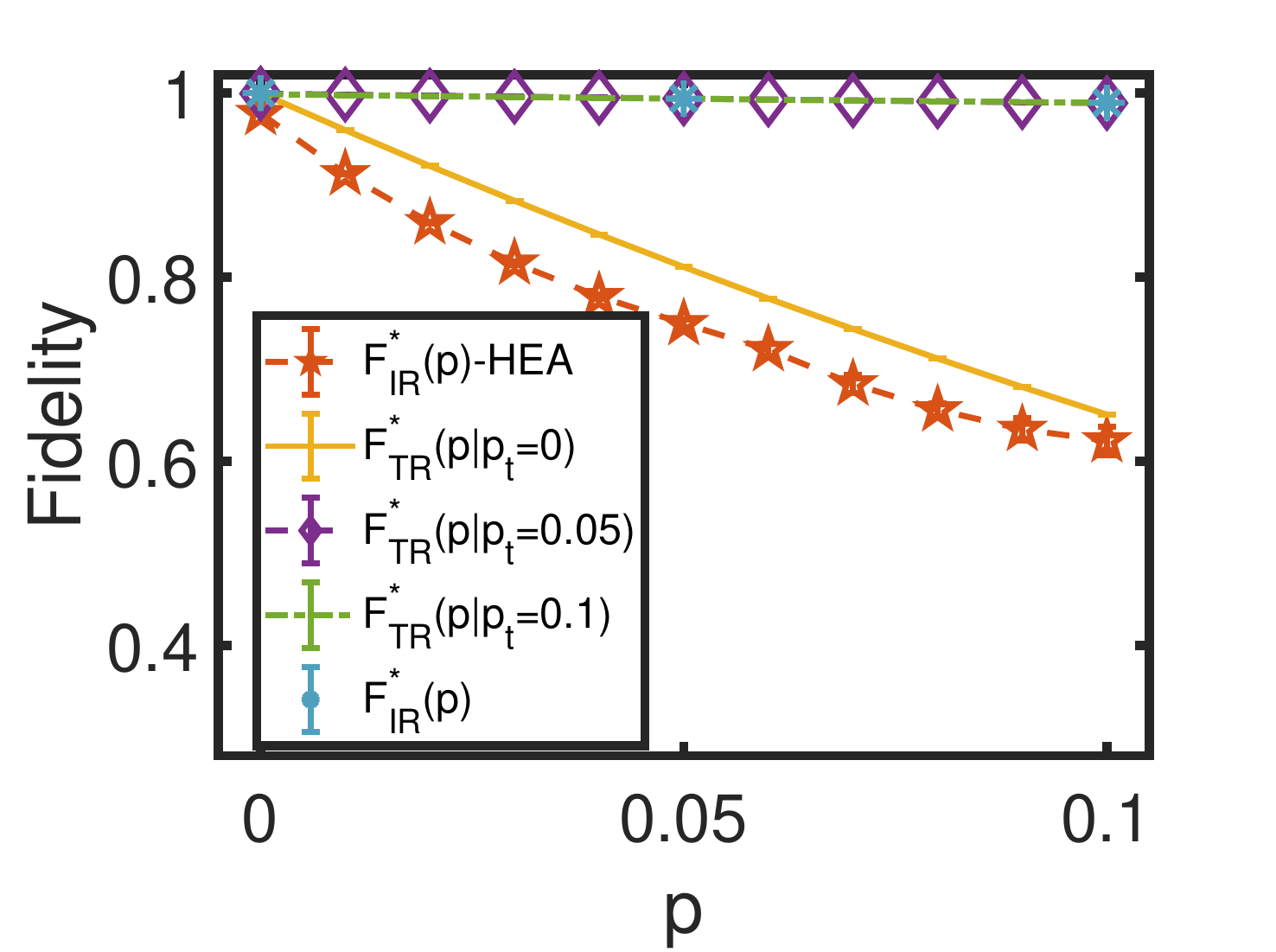}
    \end{minipage}\label{fig:p-QCEAT_b}}
    \\
     \subfigure[]{}{
    \begin{minipage}[t]{0.2\textwidth}
    \centering
    \includegraphics[width=1.25\textwidth]{Fig10_p_QCEAT_c.pdf}
    \end{minipage}\label{fig:Delta-p_c}}
    \hspace{0.2cm}
    \subfigure[]{}{
    \begin{minipage}[t]{0.2\textwidth}
    \centering
    \includegraphics[width=1.25\textwidth]{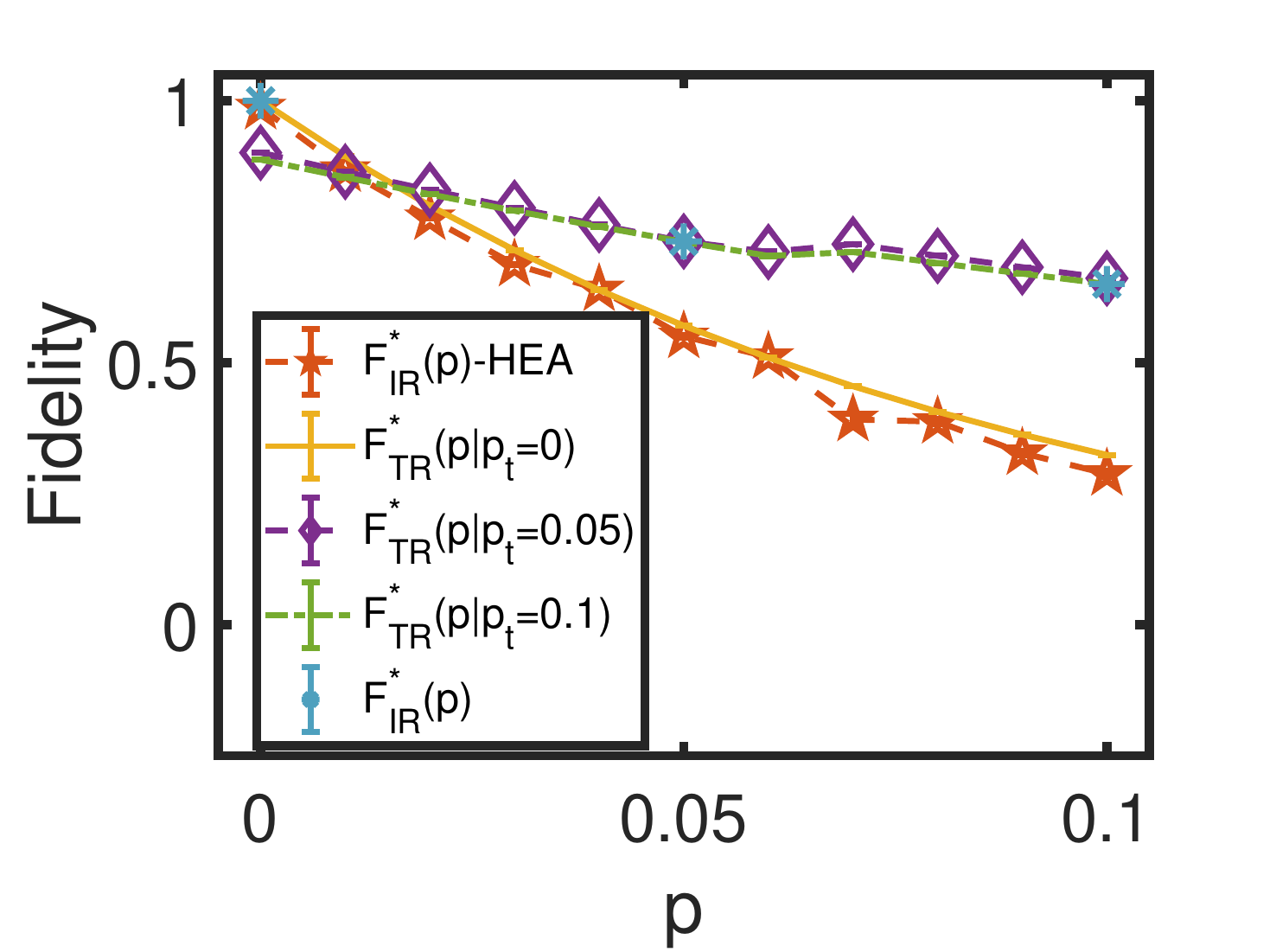}
 
    \end{minipage}\label{fig:p-QCEAT_d}}
    \caption{Comparison of the robustness of the HEA and the circuits generated by QCEAT (Fig.~\ref{fig:incoherent-circuit}) in approximating the ground energies for the hydrogen molecule (a)-(b) and the Heisenberg model (c)-(d) as the dephasing rates $p$ increases.}
   \label{fig:p-QCEAT}
\end{figure}

To see the performance of QCEAT circuits, we first discuss the imperfection-robustness scenario. In Figs.~\ref{fig:p-QCEAT}(a)-(b) we show the average energy and the fidelity as a function of $p$ for the hydrogen molecule using different circuits. In Figs.~\ref{fig:p-QCEAT}(c)-(d), the corresponding quantities for the Heisenberg Hamiltonian are presented. By using the circuit, obtained by the QCEAT algorithm at each $p$, the performance shows significant improvement over the HEA. In particular, for the hydrogen molecule the fidelity remains very close to $1$ even for $p=0.1$. 

Regarding training-robustness, we consider the three circuits obtained for $p_\text{t}=0$, $p_\text{t}=0.05$ and $p_\text{t}=0.1$ (see Fig.~\ref{fig:incoherent-circuit}) for each Hamiltonian. One can compute the average energy and the fidelity when these circuits are used on other quantum simulators with different values of $p$. The results for the average energy and the fidelity is shown in Fig.~\ref{fig:p-QCEAT}. Remarkably, while the QCEAT circuit trained with $p_\text{t}=0$ shows poor performance close to the HEA, the other circuits trained with nonzero dephasing rates provide significant improvement over HEA across the whole considered range of $p_\text{t}$. Interestingly, the circuits trained by the QCEAT with nonzero $p_\text{t}$ perform very close to each other indicating excellent training-robustness. In other words, if one designs a circuit through QCEAT training in a noisy quantum simulator, the same circuit can be used in other noisy machines with similar performance. 

\begin{table}[htp]
     \centering
    \begin{tabular}{|c|c|c|c|c|c|c|}
    \hline
    \multirow{2}*{$p_\text{t}$} &\multicolumn{2}{c|}{0} &\multicolumn{2}{c|}{0.05} & \multicolumn{2}{c|}{0.1}\\
    \cline{2-7}
     & \#R    &\#CX &\#R &\#CX &\#R & \#CX\\
    \hline
   $H_\text{h}$&4    &4 &4 &3 &4 &3\\
     \hline
    {$H_\text{AF}$}&8    &7 &9 &3 &9 &3\\
     \hline
	\end{tabular}
	\caption{The number of single- and two-qubits gates of the six circuits generated by QCEAT (Fig.~\ref{fig:incoherent-circuit}) for the hydrogen molecule $H_\text{h}$ and the Heisenberg model $H_\text{AF}$ under different dephasing rates $p$. \#$\text{R}$ represents the number of single-qubit gates and \#$\text{CX}$ represents the number of two-qubit gates.}
	\label{table:incoherent}
\end{table}

To understand the enhancement in both the imperfection- and training-robustness against incoherent noise one needs to investigate the number of CX gates in the QCEAT circuits. In Table~\ref{table:incoherent} we provide the number of gates for both the hydrogen molecule and the Heisenberg Hamiltonian for various values of $p$. By looking at the number of CX gates, one can see that the QCEAT algorithm naturally reduces the number of CX gates as $p$ becomes nonzero. This is an interesting observation as in the QCEAT algorithm we solely focus on minimizing the noisy average energy but the algorithm learns to minimize the number of CX gates naturally. Note that every CX gate in the circuit is a source of dephasing which deteriorates the performance of the circuit. This not only improves the imperfection-robustness but also enhances the training-robustness too. The reason that QCEAT can successfully finds a robust design lies in its nature. Indeed, the QCEAT algorithm starts with simple circuits and adds to their complexity as the evolution takes place. Thus, the algorithm is always capable of minimizing the complexity of the circuit.

\section{Scalability}

\begin{figure*}[htp]
	\centering
	\includegraphics[width=1.6\columnwidth]{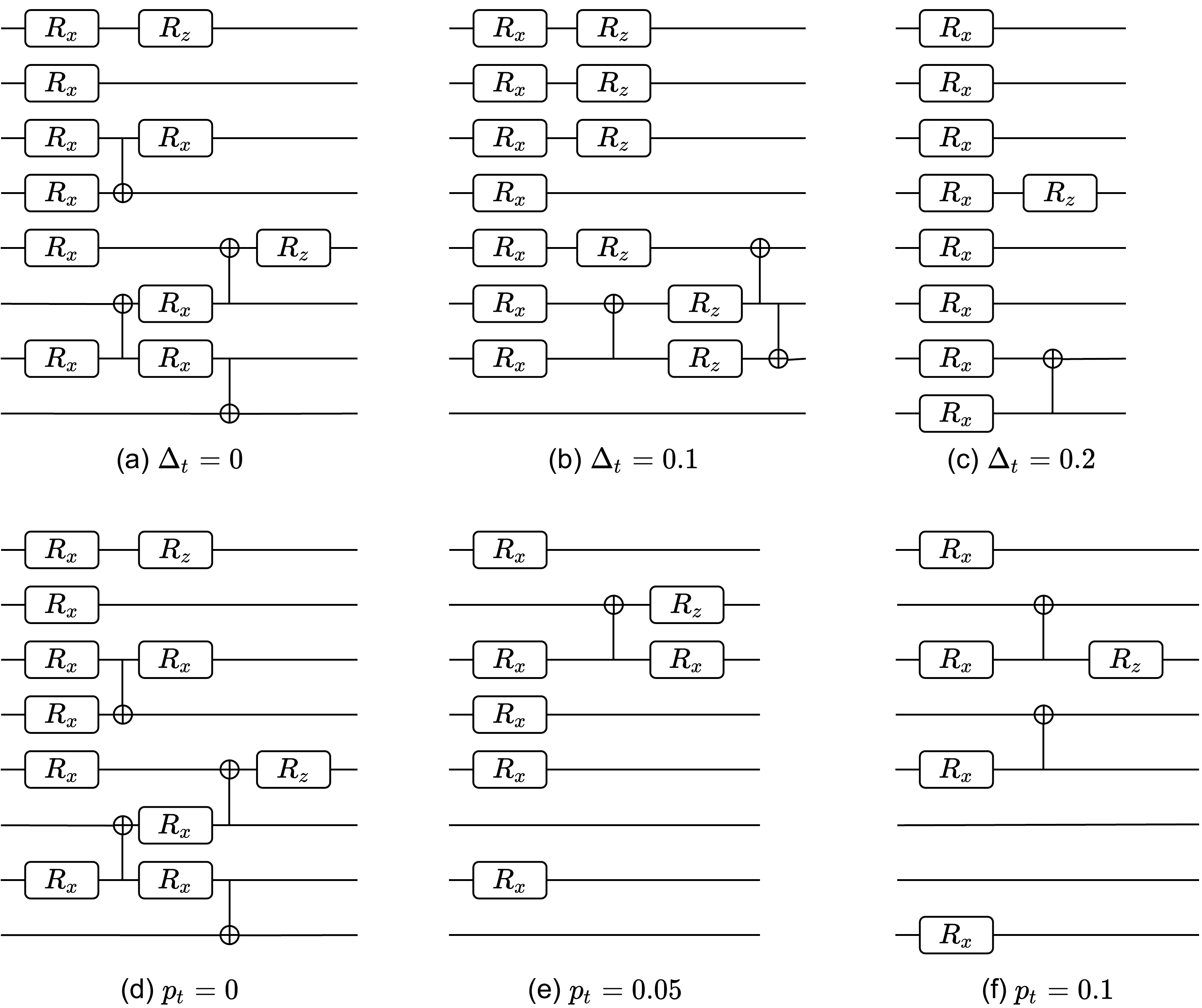}
	\caption{The optimized circuit structure generated by the QCEAT under different values of coherent noise $\Delta_\text{t}=0,0.1,0.2$ and various strengths of incoherent noise $p_\text{t}=0,0.05,0.1$, for the water molecule.
    }\label{fig:water-circuit}
\end{figure*}

A key problem for scalability of the QCEAT algorithm is to see how the population size scales with the number of qubits. The true answer to this question is highly non-trivial and beyond the scope of this paper. However, through investigating an 8-qubit Hamiltonian, we show that the population size does not necessarily need to grow proportional to the system size. In fact, we show that a population of the same size as the $N=4$ qubit case is enough to reach an optimal circuit for $N=8$. This is a remarkable observation which shows that the QCEAT algorithm can be scaled up for larger systems without demanding huge increase in the population size. To show this, we consider an $8$-qubit Hamiltonian for simulating the water molecule H$_2$O. The Hamiltonian of water molecule originals from package pyscf~\cite{sun2018pyscf}. For the sake of brevity, we do not present the Hamiltonian here as it contains many terms. The simulation of the $8$-qubit H$_2$O Hamiltonian can be performed on a HEA with two layers. This means that with HEA, one needs $14$ CX gates and $48$ parameters to optimize. The circuits generated by the QCEAT algorithm, for every $\Delta_\text{t}$ and $p_\text{t}$, are a lot more resource-efficient. In Figs.~\ref{fig:water-circuit}(a)-(f) we show six designs of the circuit, for simulating the water molecule with $\Delta_\text{t} = 0$, $\Delta_\text{t} = 0.1$, $\Delta_\text{t} = 0.2$, $p_\text{t} = 0$, $p_\text{t} = 0.05$, and $p_\text{t} = 0.1$, respectively.

\begin{figure*}[htbp]
    \centering
    \subfigcapskip=-145pt
    \subfigure[]{}{
    \begin{minipage}[t]{0.48\textwidth}
    \centering
    \includegraphics[width=1\textwidth]{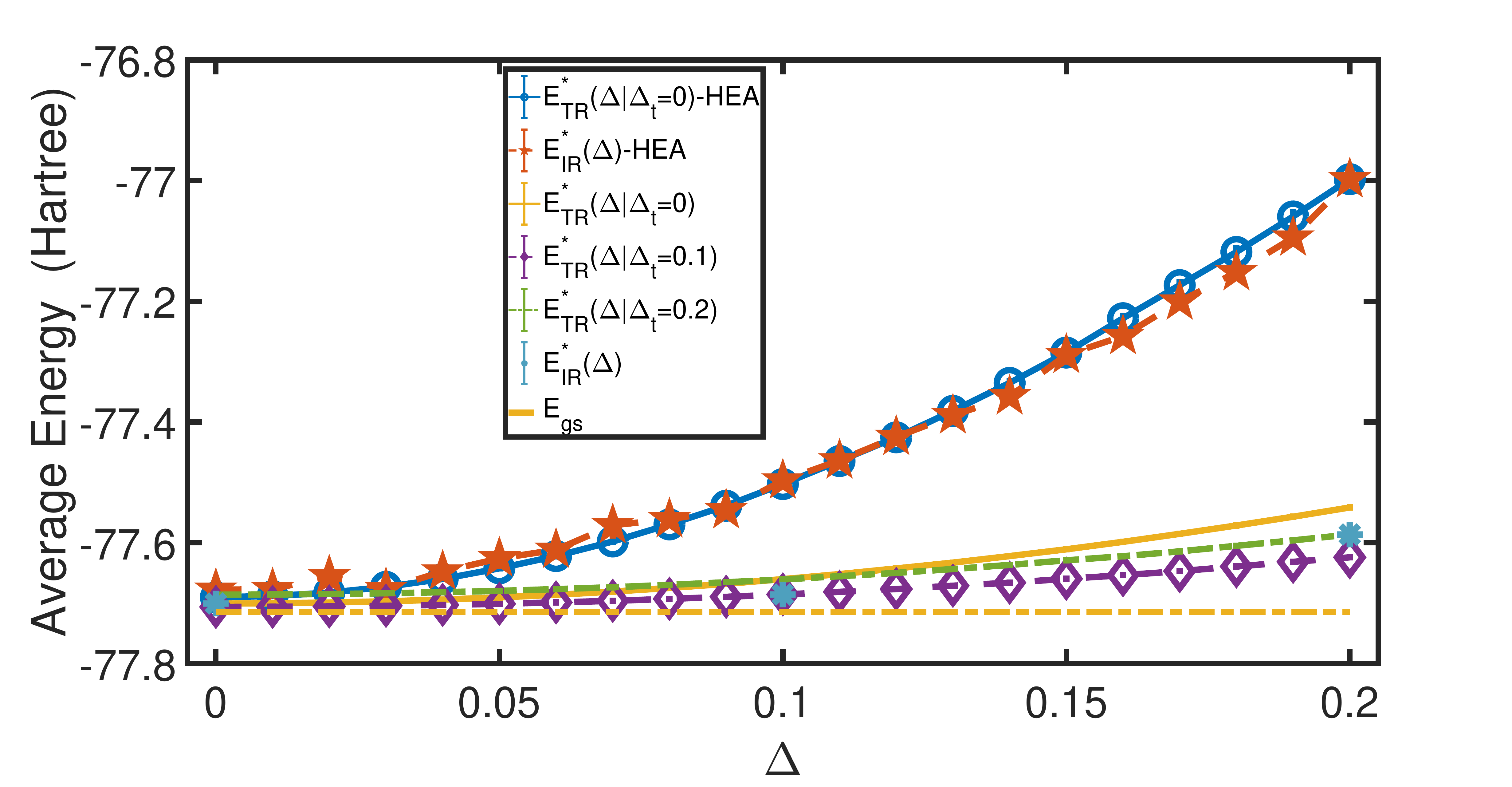}
    \end{minipage}\label{fig:water-QCEAT_a}}
    \hspace{0.2cm}
    \subfigure[]{}{
    \begin{minipage}[t]{0.48\textwidth}
    \centering
    \includegraphics[width=1\textwidth]{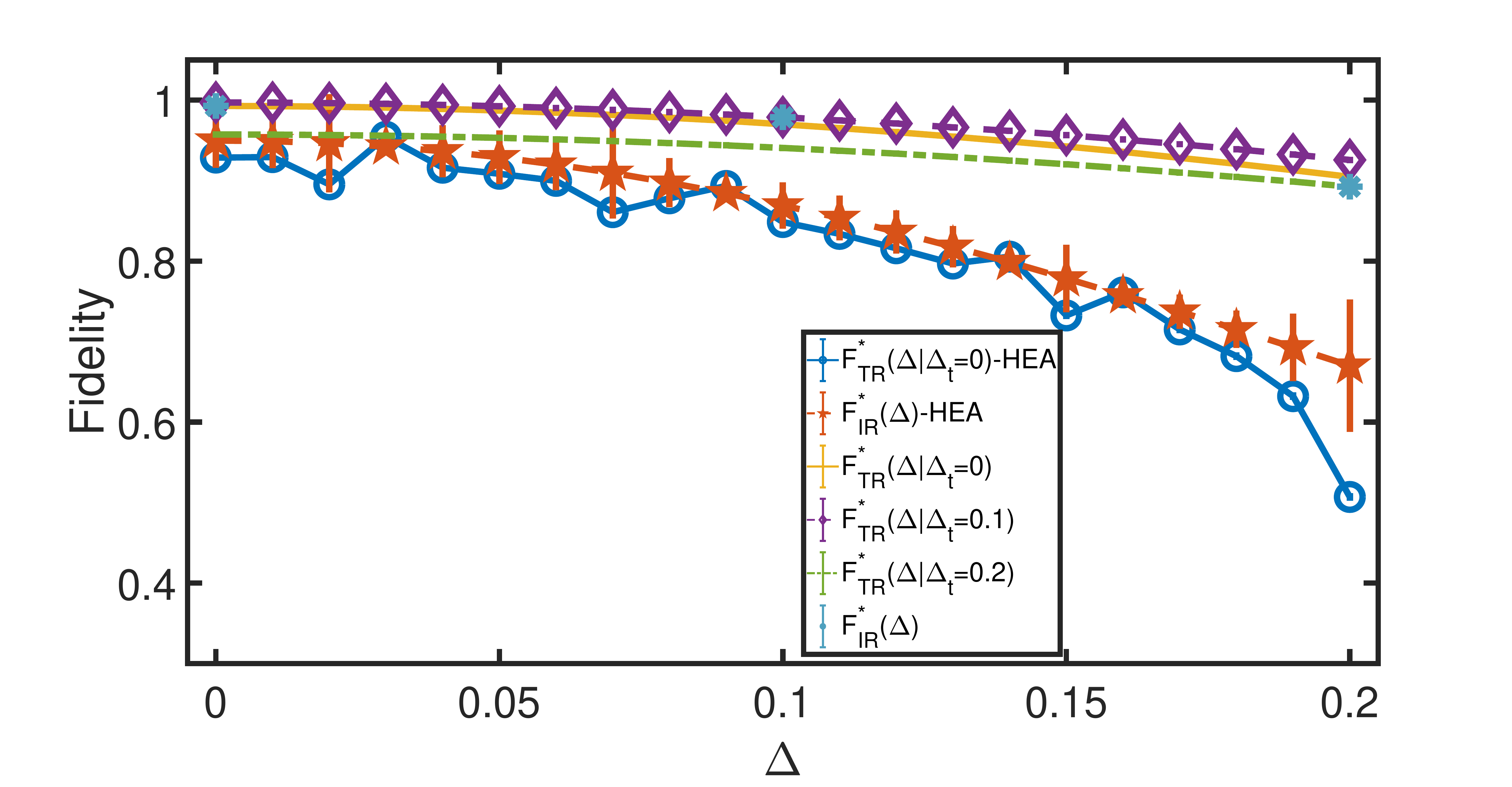}
    \end{minipage}\label{fig:water-QCEAT_b}}
    \\
     \subfigure[]{}{
    \begin{minipage}[t]{0.48\textwidth}
    \centering
    \includegraphics[width=1\textwidth]{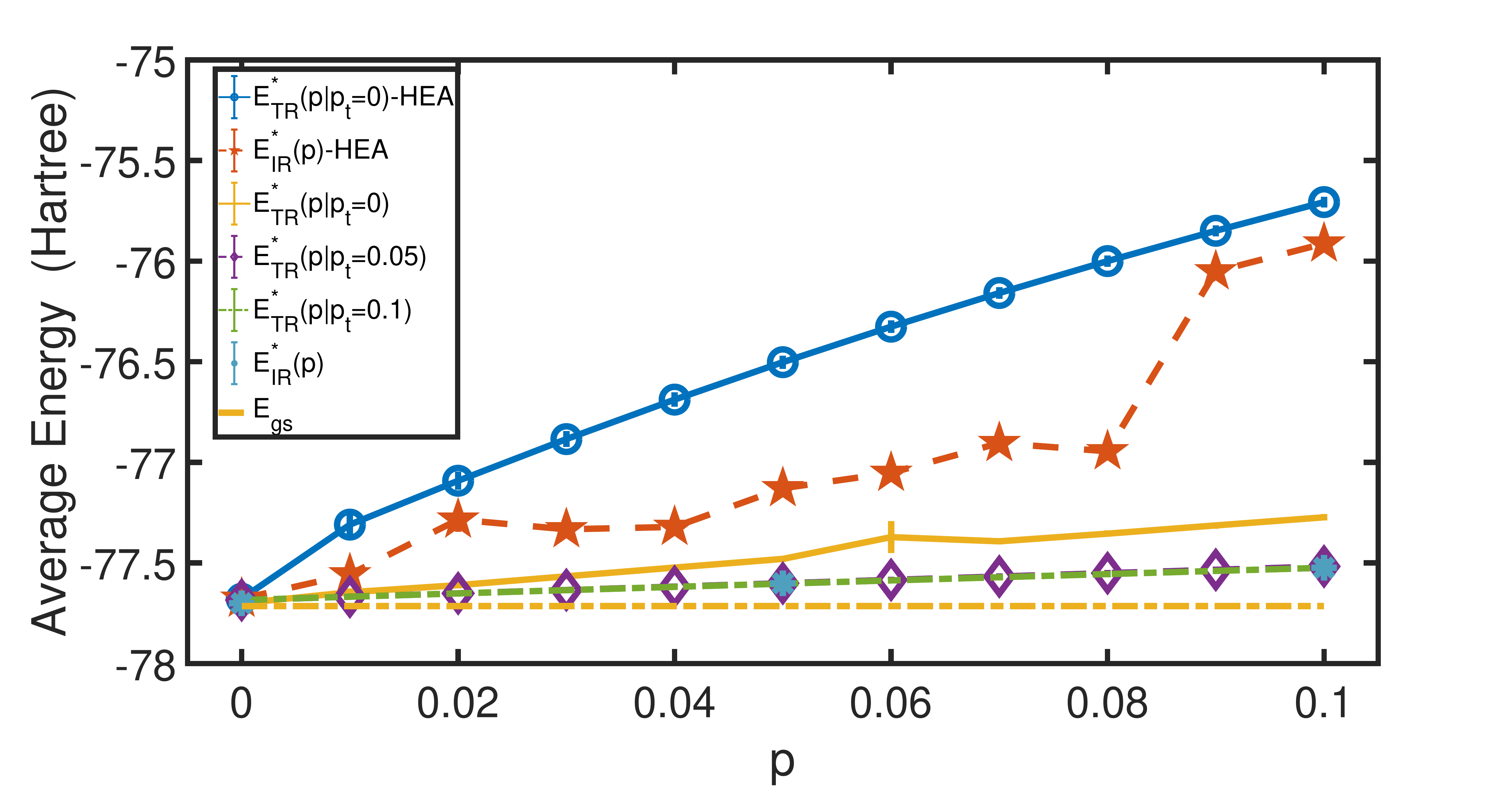}
    \end{minipage}\label{fig:water-QCEAT_c}}
    \hspace{0.2cm}
    \subfigure[]{}{
    \begin{minipage}[t]{0.48\textwidth}
    \centering
    \includegraphics[width=1\textwidth]{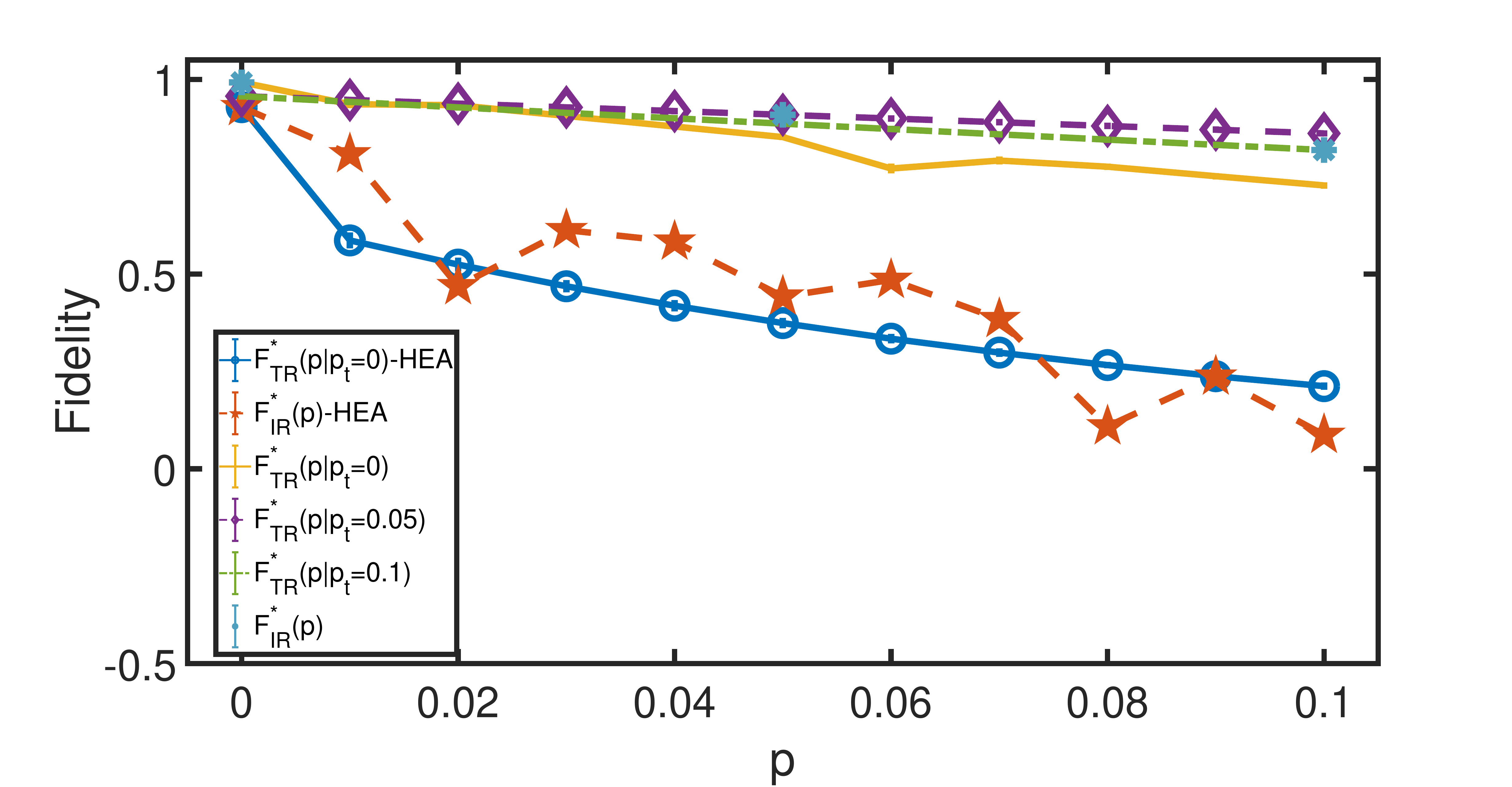}
 
    \end{minipage}\label{fig:water-QCEAT_d}}
    \caption{Comparison of the robustness of the HEA and the circuits generated by QCEAT (see Fig.~\ref{fig:water-circuit}) in approximating the ground energy for the water molecule under the action of: (a)-(b) coherent noise; and (c)-(d) the incoherent noise, respectively.
   }
   \label{fig:water-QCEAT}
\end{figure*}

In the coherent noise, to compare the imperfection- and training-robustness of the QCEAT circuits with HEA, in Figs.~\ref{fig:water-QCEAT}(a)-(b), we plot the average energy and the final fidelity as functions of $\Delta$, respectively. For every value of $\Delta_\text{t}$, a new circuit is trained. As the figures clearly show, the average energies, namely $E_{\text{IR}}^*(\Delta)$-HEA and $E_{\text{TR}}^*(\Delta|\Delta_\text{t}=0)$-HEA, and their corresponding fidelities, namely $F_{\text{IR}}^*(\Delta)$-HEA and $F_{\text{TR}}^*(\Delta|\Delta_\text{t}=0)$-HEA, for the HEA, vary drastically as $\Delta$ increases, indicating the HEA neither provides imperfection- nor training-robustness. On the other hand, the circuits designed by QCEAT show significantly higher robustness in their average energy $E_{\text{TR}}^*(\Delta|\Delta_\text{t})$ and their corresponding final fidelity $F_{\text{TR}}^*(\Delta|\Delta_\text{t})$, for various choices of $\Delta_\text{t}$.

Similarly, in Figs.~\ref{fig:water-QCEAT}(c)-(d), we plot the average energy and the final fidelity as a function of $p$ for the water molecule under the incoherent noise. For the HEA ansatz, the average energies $E_{\text{IR}}^*(p)$-HEA and $E_{\text{TR}}^*(p|p_\text{t}=0)$-HEA, and their corresponding fidelities, namely $F_{\text{IR}}^*(p)$-HEA and $F_{\text{TR}}^*(p|p_\text{t}=0)$-HEA, show very little imperfection- and training-robustness. The circuits generated by the QCEAT, however, reveal significant enhancement in both imperfection- and training-robustness for their average energy, i.e. $E_{\text{IR}}^*(p)$ and $E_{\text{TR}}^*(p|p_\text{t})$, and their corresponding final fidelities, i.e. $F_{\text{IR}}^*(p)$ and $F_{\text{TR}}^*(p|p_\text{t})$. These analyses show that even for a doubly enlarged system, the same population size can achieve very high-quality performance. Although this is not a rigorous proof, it demonstrates that the population size does not necessarily need to be increased proportionally to the system size.

\section{Conclusion}

Considering the imperfect nature of NISQ simulators, robustness against noise and training is a crucial requirement for the success of any quantum algorithm. The VQAs are among the most promising approach for achieving quantum advantage. VQE is a prominent example of such an algorithm that has been widely used for simulating the ground state of many-body systems as well as chemical structures. In this paper, we first describe two notions of robustness for quantum variational algorithms against coherent and incoherent imperfections as well as training. We show that fixed circuit designs, such as the HEA, do not provide robust circuits against noise. Therefore, we develop the QCEAT algorithm which is a versatile evolutionary algorithm for designing a quantum circuit starting from a simple structure and gradually making itself more complex without demanding any prior assumptions.
To show the generality of our results we consider different Hamiltonians for the hydrogen molecule, the water molecule and the anti-ferromagnetic Heisenberg model, and design several circuits under different types and strengths of noises.
The resulted circuits show significant imperfection- and training-robustness in comparison with the HEA. Remarkably, the QCEAT algorithm naturally minimizes the number of gates in the circuit, and not only significantly shortens the depth of the circuit but also reduces the number of parameters which need to be optimized. This means that the circuits designed by the QCEAT are also far more resource-efficient than the widely used HEA. 
Finally, we show that the population size does not necessarily grow with the number of qubits in the system, demonstrating that the QCEAT algorithm can potentially be scaled up for large systems. In the meanwhile, despite its robustness, the QCEAT algorithm is computationally expensive for larger quantum systems. Hence, as future work, we will explore how to make QCEAT work more efficiently for large quantum systems.

\section*{Acknowledgments}
The authors gratefully acknowledge the grant from National Key R\&D Program of China, Grant No.2018YFA0306703. A.B. acknowledges support from the National Science Foundation of China (Grants No. 12050410253 and No. 92065115) and the Ministry of Science and Technology of China (Grant No. QNJ2021167001L). R.-B.W. thanks the National Science Foundation of China (Grants No. 61833010 and No. 61773232), and a grant from the Institute for Guo Qiang, Tsinghua University.


%

\end{document}